\documentclass[prd,aps,eqsecnum,floatfix,nofootinbib,preprint,tightenlines]{revtex4}
\pdfoutput=1

\usepackage[T1]{fontenc}
\usepackage{bm,bbm}
\usepackage{graphics}
\usepackage{amssymb}
\usepackage{amsmath}
\usepackage{hyperref}
\usepackage{slashed}
\usepackage{mathrsfs}
\usepackage{esint}
\usepackage[normalem]{ulem}
\usepackage{longtable}
\usepackage{cancel}
\usepackage{xcolor}
\usepackage{float}
\usepackage{verbatim}
\usepackage{graphicx}
\usepackage[small]{subfigure}

\newcommand{\dd}{\mathrm d}

\definecolor{purple}{rgb}{0.57, 0.36, 0.51}

\newcommand{\FOPT}{\rm{FOPT}}
\newcommand{\CIPT}{\rm{CIPT}}
\definecolor{darkblue}{rgb}{0.0, 0.0, 0.55}
\definecolor{darkmidnightblue}{rgb}{0.0, 0.2, 0.4}

\usepackage{amsthm}
\newtheorem{definition}{Definition}[section]
\newtheorem{theorem}[definition]{Theorem}
\newtheorem{corollary}{Corollary}[definition]

\begin{document}

\begin{flushright}
UWThPh-2023-14\\
\end{flushright}
\vspace{0.5cm}

\begin{center}
\begin{boldmath}
{\large{\bf Mathematical Aspects of the Asymptotic Expansion in Contour Improved Perturbation Theory for Hadronic Tau Decays}}\\[8mm]
\end{boldmath}
Nestor G. Gracia,$^{a,b}$ Andr\'e H. Hoang,$^{b,c}$ and Vicent Mateu$^{a}$\\[8 mm]
$^a$Departamento de F\'isica Fundamental e IUFFyM, Universidad de Salamanca,\\E-37008 Salamanca, Spain\\
[5mm]
$^b$Faculty of Physics, University of Vienna, Boltzmanngasse 5, A-1090 Wien, Austria
\\[10mm]
\end{center}

\begin{quotation}
Recently, it was demonstrated that the discrepancy between the fixed-order and contour-improved (CIPT) perturbative expansions for $\tau$-lepton decay hadronic spectral function moments, which had been affecting the precision of $\alpha_s$ determinations for many years, is related to the CIPT expansion being inconsistent with the standard formulation of the operator product expansion. Even though the problem can be alleviated phenomenologically for the most part by employing a renormalon-free scheme for the gluon-condensate matrix element, the principal inconsistency of CIPT remains. The CIPT expansion is special because it is not a power expansion, but represents an asymptotic expansion in a sequence of functions of the strong coupling. In this article we provide a closer look at the mathematical aspects of the asymptotic sequence of the functions the CIPT method is based on, and we expose the origin of the CIPT inconsistency as well as the reasons for its apparent good convergence at low orders. Our results are of general interest, and may in particular provide a useful tool to check for the consistency of expansion methods that are similar to CIPT.
\end{quotation}

\section{Introduction}
\label{sec:intro}

The comparison of moments of the inclusive hadronic $\tau$-decay invariant-mass spectral distribution obtained at LEP~\cite{Davier:2013sfa} with the corresponding theoretical predictions, based on finite-energy sum rules involving the Adler function, is one of the most precise methods to determine the QCD strong coupling $\alpha_s$ at the scale of the $\tau$ lepton mass~\cite{ParticleDataGroup:2022pth}. The theoretical predictions for these finite-energy sum rules involve perturbative series for weighted contour integrals over the invariant mass of the vacuum polarization function that encode the main dependence of the moments on $\alpha_s$.
For many years a systematic theoretical discrepancy has persisted for these perturbative series~\cite{Proceedings:2011zvx,dEnterria:2022hzv} related to two renormalization scale-setting prescriptions, called fixed-order perturbation theory (FOPT) and contour-improved perturbation theory (CIPT)~\cite{Pivovarov:1991rh,LeDiberder:1992jjr}. While FOPT represents a simple expansion in powers of the strong coupling at a certain renormalization scale, CIPT is based on integrations over the strong coupling renormalization scale and is therefore an expansion in nontrivial functions of the strong coupling.
In moments for which the leading nonperturbative correction, coming from the dimension-four gluon-condensate (GC) matrix element in the operator production expansion (OPE), becomes suppressed due to the integration with the corresponding weight function (which are the most relevant for $\alpha_s$ determinations) the CIPT scale setting leads to systematically smaller values for the truncated perturbative series, resulting in larger values for the extracted strong coupling.

It was claimed already some time ago in Refs.~\cite{Beneke:2008ad,Beneke:2012vb} based on the study of renormalon models of the Adler function, that the discrepancy between CIPT and FOPT is due to an inconsistency of the CIPT expansion. Recently, in Refs.~\cite{Hoang:2020mkw,Hoang:2021nlz}, it was shown that for these moments the CIPT expansion, in contrast to FOPT, does not lead to a corresponding suppression of the moment's quartic sensitivity to infrared (IR) momenta, and the associated suppression of the infrared renormalon does not take place. This renders CIPT, as a matter of principle, inconsistent with the standard formulation of the OPE, even though the CIPT series typically exhibits an apparently excellent behavior at low orders.
As far as the hadronic $\tau$ decay spectral-function moments are concerned, where this issue is numerically dominated by the dimension-four GC renormalon~\cite{Hoang:2020mkw,Hoang:2021nlz}, the CIPT problem can be alleviated phenomenologically to a large extent by employing a renormalon-free scheme for the GC matrix element~\cite{Benitez-Rathgeb:2022yqb,Benitez-Rathgeb:2022hfj}.
Even though the CIPT problem for hadronic $\tau$-decay spectral function moments can therefore be considered as resolved for phenomenological purposes and a quantitative description of the CIPT inconsistency (called the ``asymptotic separation'') has been provided in Refs.~\cite{Hoang:2020mkw,Hoang:2021nlz}, the deeper mathematical background of why CIPT turns out to be inconsistent with the OPE has not yet been studied.

It is the purpose of this article to provide such a mathematical examination. An important motivation for our study is that, once the mathematical origin of the problem is specified in a way independent of the application to the $\tau$ hadronic spectral function moments, the consistency of other perturbative expansion methods where integrations over the strong-coupling renormalization scale are employed can be checked. For the most part of our study we use the leading logarithmic (one-loop) approximation for the strong coupling evolution and the large-$\beta_0$ approximation, where essentially at all stages of our analysis we can rely on fully analytical results. We also check by numerical analyses that all our conclusions are (beyond any reasonable doubt) valid as well in full QCD. Our main finding can be summarized in the following simple statement: {\bf ``A convergent series in FOPT in general leads to a divergent series in CIPT''}. We provide the general mathematical criteria on which this statement is based, and also clarify why CIPT frequently appears to provide a more convergent series expansion than FOPT at low orders.

The article is organized as follows: In Sec.~\ref{sec:W1largeb0} we set up our notation and prove that the set of expansion functions of CIPT indeed form well-defined asymptotic sequences, which ensures that CIPT provides well-defined asymptotic expansions. However, these asymptotic sequences lack an important type of uniformity property due to zeros for coupling values that become ever smaller at large orders. Here we also prove analytically that a generic factorially divergent series contained in the Adler function (related to an infrared renormalon), which yields a FOPT series with a finite radius of convergence when the corresponding OPE correction vanishes, leads to a divergent CIPT series for any value of the strong coupling. This fact was already discussed in Refs.~\cite{Hoang:2020mkw,Hoang:2021nlz,Benitez-Rathgeb:2022yqb} based on phenomenological finite-order studies, but not proved rigorously based on the all-order behavior of the actual series. In Sec.~\ref{sec:W1Deeper} we discuss the reexpansion of CIPT series in terms of FOPT and vice versa. We demonstrate that, while the functions defining the asymptotic expansion of CIPT can be represented by FOPT series with a finite radius of convergence, the reverse is not true. In fact, the asymptotic expansion of a single power of the strong coupling in terms of the CIPT expansion functions is divergent for any value of $\alpha_s$. We provide arguments supporting that this property is related to the nonuniformity and the zeros of the CIPT expansion functions. This finding is the basis of the boldfaced statement above and the central result of this work. All these results are based on explicit analytic expressions derived using the leading logarithmic strong coupling evolution and the large-$\beta_0$ approximation. Finally, in Sec.~\ref{sec:WxellQCD} we provide numerical evidence that these findings also apply in full QCD, where explicit all-order analytical expressions are not available. In Sec.~\ref{sec:conclusions} we conclude. Lastly, we also add Appendix~\ref{sec:theorem}, where we state a number of mathematical definitions, theorems, and corollaries that we use in the main body of the article concerning the convergence properties of series, series of functions and the concept of asymptotic expansions.
Even though we assume that they are familiar to many researchers, we quote them in the article explicitly for completeness, since some of them, such as the Weierstrass double series theorem, are rarely used in high-energy physics applications.
In particular we quote the definition for so-called asymptotic sequences which generalize the concept of asymptotic power expansions. This generalization is relevant
because CIPT is not an expansion in powers of the strong coupling but in more complicated functions of the strong coupling with a nontrivial analytic structure. We recommend the reader not familiar with this general definition of asymptotic expansions or the Weierstrass theorem to start the article with the Appendix~\ref{sec:theorem}.

\section{Spectral Function Moment Series in the Large-$\beta_0$ Approximation}
\label{sec:W1largeb0}

\subsection{Notation and Conventions}
\label{sec:notation}

We write the perturbative series for the reduced Adler function for strange plus nonstrange light-quark production in the form\footnote{The coefficients $\bar c_{n ,k}$ are defined in the perturbative expansion of the vacuum polarization function, which reads
$\Pi (s, \mu^2) = - \frac{1}{4 \pi^2} [ L + \sum_{n = 1} a^i(\mu^2) \sum_{k = 0}^n \bar c_{n ,k} L^k ]$,
with $L=\ln( - s/\mu^2)$. The reduced Adler function is defined by the relation $ \hat D(s) = - 4 \pi^2 s \frac{\dd \Pi (s)}{\dd s}- 1$ and is renormalization scale invariant.}
\begin{equation}
\label{eq:AdlerD}
\hat D(s) =
\sum_{n=1}^\infty \, \bar c_{n,1} \,a^n(-s)=\sum_{n=1}^\infty a^n \sum_{k=1}^{n}\, k\,\bar c_{n,k}\ln^{k-1}(-x)\,,
\end{equation}
where
we define ($\beta_0=11-2n_f/3$ with $n_f=3$)
\begin{equation}
\label{eq:adef}
a(\mu^2)\equiv \frac{\beta_0\alpha_s(\mu^2)}{4\pi}\,,
\end{equation}
for the rescaled strong coupling.
In Eq.~\eqref{eq:AdlerD} and in what follows we use the shorthand notation $a\equiv a(s_0)$ and $x = s/s_0$, where frequently in phenomenological applications one has $s_0=m_\tau^2$.
The series for the $\tau$ hadronic spectral function moments which are derived from the Adler function are defined by the contour integrals
\begin{equation}\label{eq:moments}
\delta^{(0)}_{W(x)}(a)=\frac{1}{2i\pi}\oint_{\vert x\vert=1}\!\frac{\dd x}{x}W(x)\hat{D}(s_0x)\,,
\end{equation}
where the contour starts/ends at $x=1\pm i 0$ (or $s=s_0\pm i 0$) and is a counterclockwise path in the complex $x$-plane (or $s$-plane) around the origin~\cite{Braaten:1991qm,LeDiberder:1992jjr}.
The weight function $W(x)$ is a polynomial with the property $W(1)=0$ for physical applications.
Note that we suppress the dependence of the moments on the physical scale $s_0$ since it comes solely from the argument of $a=a(s_0)$.
In the following we also frequently use the shorthand notation $\delta^{(0)}_{\ell}$ instead of $\delta^{(0)}_{W(x)}$ when considering the simple monomial weight function $W(x)=(-x)^\ell$.

The CIPT expansion for $\delta^{(0)}_{W(x)}$ starts from the perturbative series for $\hat D(s)$ given in powers of $a(-s)$, and the contour integral
is done over powers of the complex-valued strong coupling $a(-s)=a(-x s_0)$. This yields
\begin{equation}
\label{eq:deltaCIPTdef}
\delta_{{\CIPT}, \ell}^{(0)}(a)=\sum_{n=1}^\infty
\bar{c}_{n,1} H_{n, \ell}(a)\,,
\end{equation}
where
\begin{equation}
\label{Hndef}
H_{n, \ell}(a) \equiv \frac{1}{2 i \pi}\oint_{|x|=1}\!
\frac{\dd x}{x} (-x)^\ell a^n(-xs_0)\,.
\end{equation}
The resulting series is an asymptotic expansion in terms of the functions $H_{n, \ell}(a)$ with fixed $\ell$ depending on the reference strong coupling $a$, which is the quantity determined from the comparison to experimental data in strong coupling determinations.
The FOPT expansion for $\delta^{(0)}_{W(x)}$ starts from the perturbative series for $\hat D(s)$ given in powers of $a=a(s_0)$.
Complex phases then appear from the powers of $\ln(-x)$ in the coefficients of this series, and the contour integral is computed over
the polynomials in $\ln(-x)$. This yields
\begin{equation}
\label{eq:deltaFOPT}
\delta_{{\FOPT}, \ell}^{(0)}(a)=\sum_{n=1}^\infty d^{\text{FOPT}}_{n, \ell} a^n\,,
\quad d^{\text{FOPT}}_{n, \ell}=
\sum_{k=1}^n k\,\bar{c}_{n, k} I_{k-1, \ell}\,,
\end{equation}
where
\begin{equation}
\label{eq:Ikdef}
I_{k, \ell}\equiv \frac{1}{2i\pi}\oint_{\vert x\vert=1} \!\frac{\dd x}{x}(-x)^\ell\ln^k(-x)\,.
\end{equation}
The integrals $I_{k,\ell}$ can be determined analytically and read:
\begin{align}
\label{eq:InlIntegrals}
I_{k\neq 0, \ell\neq 0} &= \frac{\Gamma(k+1,i\pi\ell,-i\pi\ell)}{2i\pi(-\ell)^{k+1}}
=
-\frac{k!(-1)^\ell}{2(-\ell)^k} \sum_{j = 0}^{k - 1} \frac{(i \ell \pi)^j \bigl[1+(-1)^j\bigr]}{(j + 1)!}\\
& = \pi^k (- 1)^{\ell + k} \sum_{n = 0}^{\infty}
\frac{(\pi \ell)^n}{(k + 1)_{n+1}}\cos \Bigl[ \frac{\pi}{2} (k + n) \Bigr]\,,\nonumber\\
\label{eq:InlIntegrals0}
I_{k,0}&=\frac{(i\pi)^k\bigl[1+(-1)^k\bigr]}{2(k+1)}=
\frac{\pi^k(-1)^k}{k+1}\cos\biggl(\frac{\pi k}{2}\biggr),
\end{align}
where $\Gamma(n,a,b)\equiv\Gamma(n,a)-\Gamma(n,b)$ and $I_{0,\ell} = \delta_{\ell,0}$. Note that the second equality of Eq.~\eqref{eq:InlIntegrals} stems from Eq.~\eqref{eq:incGammarel} and the third from the identity in Eq.~\eqref{eq:GammaAsy2}.
The notation $(b)_{k}$ stands for the Pochhammer symbol $(b)_{k}=\Gamma(b+k)/\Gamma(b)$.
Note that for $\ell=0$ only the leading term of the sum over $n$ in the second line survives, yielding the expression for $I_{k,0}$ given in the last line. The series $\delta_{{\FOPT}, \ell}^{(0)}$
is a common power expansion in $a$. Once the CIPT and FOPT moment series are determined, it is not possible any more to change between the FOPT and CIPT expansions through a change of renormalization scale, since the CIPT series represents a coherent analytic weighted combination of complex-valued strong coupling series with different renormalization scales. It is this difference which is at the core of our investigations.

In the large-$\beta_0$ approximation the QCD $\beta$-function for the evolution of $a(\mu^2)$ has the simple leading logarithmic form $\dd a(\mu^2)/\dd \ln \mu^2=-a^2(\mu^2)$ yielding
\begin{equation}
\label{eq:alphab0}
a(\mu^2) = \frac{a(\mu_0^2)}{1+a(\mu_0^2)\ln\Bigl(\frac{\mu^2}{\mu_0^2}\Bigr)} \,,
\end{equation}
as the relation of the coupling at different scales. In the large-$\beta_0$ approximation the Adler function is determined from single-gluon exchange diagrams dressed with infinitely many insertions of one-loop massless-quark vacuum polarization bubbles with $n_f$ flavors. Subsequently the replacement $n_f\to -3\beta_0/2$ is imposed to yield an approximation for the full QCD results including gluonic corrections. This approximation provides the correct ${\cal O}(\alpha_s)$ next-to-leading order (NLO) QCD corrections and is known to have many qualitative features of full QCD concerning corrections beyond NLO.
It has the advantage that essentially all calculations can be carried out analytically to all orders. It is this property which we rely on in most of the following sections of this article. Still, it is essential to reconfirm that the qualitative insights gained in the large-$\beta_0$ approximation also apply in full QCD. This is what we address in Sec.~\ref{sec:WxellQCD}.

In the large-$\beta_0$ approximation the all-order perturbative series for the Adler function can be conveniently written down in closed form~\cite{Broadhurst:1992si}
\begin{equation}
\label{eq:invBorelD}
\hat D(s) = \int_0^\infty \!\! \dd u \bigl[
B(u)\bigr]_{\rm Taylor} e^{-\frac{u}{a(-s)}}
= \int_0^\infty \! \dd u \Bigl[
B(u) e^{-u\ln(-x)}\Bigr]_{\rm Taylor} e^{-\frac{u}{a}}\,,
\end{equation}
where the Borel functions in the brackets
need to be Taylor expanded in powers of $u$ and the relation
$\int_0^\infty \dd u \,u^{n-1}\,e^{-\frac{u}{x}} = \Gamma(n)\,x^n$ is used. The two equalities yield the two expansions in Eq.~\eqref{eq:AdlerD} in powers of $a(-s)$ or $a$. The function $B(u)$ reads~\cite{Broadhurst:1992si}
\begin{align}
\label{eq:AdlerBorelb0}
B(u) & = \frac{128}{3\beta_0}\,\frac{e^\frac{5u}{3}}{2-u}\,\sum_{k=2}^\infty \, \frac{(-1)^k \,k}{[k^2-(1-u)^2]^2} \\
&=
\frac{8}{3\beta_0}\,\frac{e^\frac{5u}{3}}{(2-u)(u-1)}\biggl[\psi^{(1)}\!\biggl(\frac{3}{2}-\frac{u}{2}\biggr)-
\psi^{(1)}\!\biggl(2-\frac{u}{2}\biggr)+\psi^{(1)}\!\biggl(\frac{u}{2}+1\biggr)-
\psi^{(1)}\!\biggl(\frac{u}{2}+\frac{1}{2}\biggr)\!\biggr] \nonumber \\
& =
\frac{128}{3\beta_0}\,e^\frac{5u}{3}\biggl\{\frac{3}{16(2-u)} +\!\sum\limits_{p=3}^\infty \biggl[ \frac{d_2(p)}{(p-u)^2} - \frac{d_1(p)}{p-u} \biggr]
-\!\sum\limits_{p=-1}^{-\infty} \biggl[ \frac{d_2(p)}{(u-p)^2} + \frac{d_1(p)}{u-p} \biggr]\!
\biggr\},\nonumber
\end{align}
where $\psi ^{(1)}(x)=\frac{{\rm d}^2}{{\rm d}x^2}\ln[\Gamma(x)]$ is the first-order polygamma function and $d_2(p)=\frac{(-1)^p}{4(p-1)(p-2)}$, $d_1(p)=\frac{(-1)^p(3-2p)}{4(p-1)^2(p-2)^2}$. Each (single or double) pole at $u=p$ along the {\it positive real} $u$ axis corresponds to an equal-sign factorially diverging asymptotic series contribution in the coefficients of the Adler function, indicating a sensitivity of the Adler function to IR momenta $\Lambda$ of ${\cal O}(\Lambda^{2p})$. These poles (as well as the corresponding factorially diverging series contributions) are called ``IR renormalons''. Each such IR renormalon contribution has a one-to-one association with a higher dimensional OPE correction term $\sim\langle{\cal O}_p\rangle/s_0^p$, which effectively compensates for the resulting ambiguity of the perturbation series. Here, $\langle{\cal O}_p\rangle$ stands for a dimension $2p$ nonperturbative low-energy QCD matrix element that cannot be determined with perturbative methods~\cite{tHooft:1977xjm,Mueller:1984vh,Beneke:1998ui}.

For a moment with weight function $W(x)=(-x)^\ell$ all OPE matrix-element corrections of the form $\mbox{const.}\times\langle{\cal O}_p\rangle/s_0^p$ for $p\neq\ell$ are eliminated by the contour integration as can be seen from the residue theorem. As a consequence, perturbative expansion methods consistent with the OPE must yield a convergent series, at least within some region of $a$, for those terms in the Borel function $B(u)$ with a single pole renormalon of the form $1/(p-u)$~\cite{tHooft:1977xjm,Mueller:1984vh,Beneke:1998ui}.
In the following two subsections we prove analytically that FOPT satisfies this requirement, while CIPT does not. Therefore, we consider the moment $\delta_{\ell}^{(0)}$ in the FOPT and the CIPT expansions arising from the term $B_p(u) = 1/(p-u)$ with $p=2,3,\ldots$, which yields the coefficients
\begin{equation}
\label{eq:cnkb0}
\bar{c}_{n, k, p}=\frac{(-1)^{k+1}\Gamma(n)}{p^{n-k+1}\Gamma(k+1)}\,,
\end{equation}
in the Adler function series of Eq.~\eqref{eq:AdlerD}.

\subsection{FOPT Series}
\label{sec:W1FOPT}
\begin{figure*}[t]
\subfigure[]{\includegraphics[width=.45\linewidth]{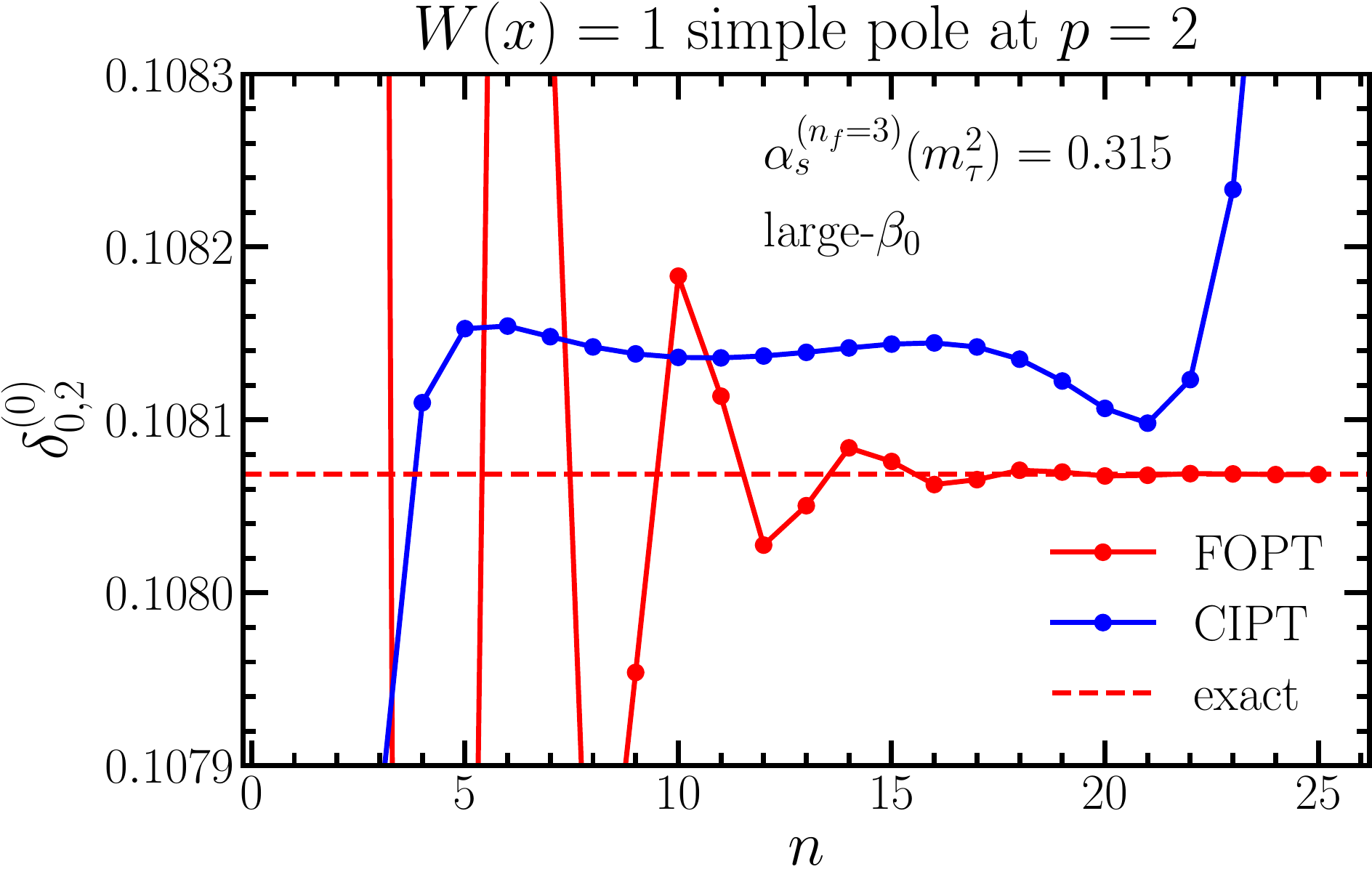}}
\subfigure[]{\includegraphics[width=.45\linewidth]{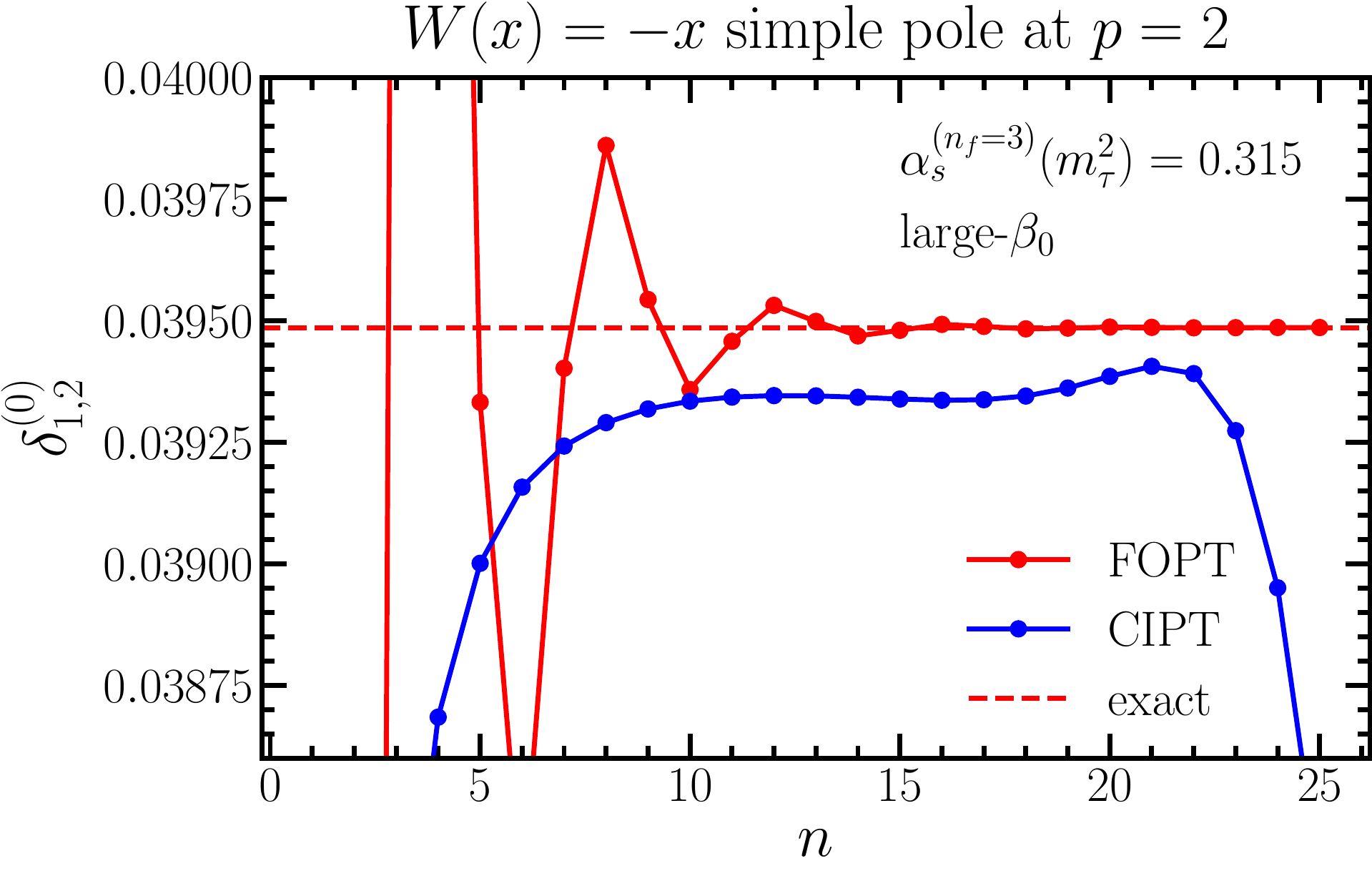}}
\caption{Partial sums in FOPT (red dots) and CIPT (blue dots) for the spectral function moments $\delta_{\ell, p}^{(0)}$ defined in Eq.~\eqref{eq:moments} using monomial weight functions
of type $(-x)^\ell$ with $\ell = 0$ (left panel) and $\ell = 1$ (right panel), including up to $n$ terms. The results assume the large-$\beta_0$ approximation and are obtained from the Adler function coefficients $\bar{c}_{n, k, 2}$ given in Eq.~(\ref{eq:cnkb0}). The horizontal, red, dashed lines indicate the value the series converges to. For our numerics, $\alpha^{(n_f=3)}_s(m_\tau^2)=0.315$ ($a=0.2256$) is employed.}
\label{fig:FOPTvsCIPT}
\end{figure*}

The contribution of the single pole $1/(p-u)$ renormalon to the FOPT moment series coefficients $d^{\text{FOPT}}_{n, \ell,p}$, defined using the coefficients of Eq.~\eqref{eq:cnkb0} in Eq.~\eqref{eq:deltaFOPT},
can be written down immediately from the expressions given above.
We found closed analytic expressions for these coefficients, simply carrying out the sums in the rightmost expression of Eq.~\eqref{eq:deltaFOPT},
for any $\ell\in\mathbb{N}_0$ and $p\in\mathbb{N}$:
\begin{align}
\label{eq:dFOPT}
d^{\text{FOPT}}_{n, 0, p} &= \frac{(- 1)^p \Gamma (n + 1, i \pi p, - i \pi
p)}{2 i \pi n p^{n + 1}},\\ \nonumber
d^{\text{FOPT}}_{n, \ell > 0, p \neq \ell} &= \frac{\ell^{- n}
\Gamma (n, i \pi \ell, -i \pi \ell) - (- 1)^{\ell + p} p^{- n} \Gamma (n,
i \pi p, -i \pi p)}{2 i \pi (\ell-p)},\\
d^{\text{FOPT}}_{n, p, p} &= \frac{\Gamma (n, i \pi p) +
\Gamma (n, - i \pi p)}{2 p^n} + \frac{\Gamma (n + 1, - i \pi p, i \pi p)}{2
\pi i p^{n + 1}}\,. \nonumber
\end{align}
Note that we have $d^{\rm FOPT}_{1, \ell>0, p}=0$.
Using the expansion of the incomplete gamma function $\Gamma(n,z)$ for $|n|\to\infty$ quoted in Sec.~\ref{sec:asymptoticformulae} we can obtain the large-$n$ asymptotic expression for the coefficients $d^{\text{FOPT}}_{n,\ell,p}$ which we can use to apply the root test according to Theorem~\ref{th:root}. From Eqs.~\eqref{eq:GammaAsy1} and (\ref{eq:GammaAsy2}) we find
\begin{equation}
\label{eq:dFOPTAsy}
d_{n, \ell, p}^{\rm FOPT} -\frac{\Gamma (n)}{p^n} \delta_{\ell,p} =
(- 1)^{\ell+1} \pi^n\! \sum_{k = 1}^\infty \frac{ (p\pi)^{k - 1}}{(n)_{k + 1}} \sin \biggl[ \frac{\pi}{2}
(k + n) \biggr] \sum_{j = 0}^{k - 1} \biggl(\frac{\ell}{p}\biggr)^j ,
\end{equation}
where the sum over $k$ on the RHS refers to the $k$-th leading term as $n\to\infty$ when $p\neq\ell$. This sum is actually absolutely convergent.
It is conspicuous that there is a universal formula for the (asymptotic) expansion in the cases $p=\ell$ and $p\neq \ell$ as $n\to\infty$. Furthermore,
for $p\neq\ell$ the leading large-$n$ expression (for $k=1$) does not depend on $p$, and its only dependence on $\ell$ is through the factor $(-1)^\ell$. The latter property implies that for $p\neq\ell$ the leading term as $n\to\infty$ in Eq.~\eqref{eq:dFOPTAsy} always cancels for physical weight functions which have the property $W(1)=0$. Interestingly, also the first $j$ subleading terms
cancel if also the first $j$ derivatives of $W(x)$ vanish at $x=1$ ($W^\prime(1)=W^{\prime\prime}(1)=\cdots= W^{(j)}(1)=0$), i.e., if $W$ is $j$-fold pinched. So for the kinematic weight function \mbox{$W_{\tau}(x)=(1-x)^3(1+x)=1-2x+2x^3-x^4$}, which is relevant for the inclusive $\tau$ hadronic decay width, we have $W(1)=W^\prime(1)=W^{\prime\prime}(1)=0$. Here the terms for $k=1,2,3$ shown in Eq.~\eqref{eq:dFOPTAsy} cancel for all $p\neq 3,4$. We note that the case $p=1$ does not arise in the Adler function, see Eq.~\eqref{eq:AdlerBorelb0}.
It is now straightforward to determine the limit superior that we need for the root convergence criterion,
\begin{align}
\label{eq:limsupellp}
\underset{n \rightarrow \infty}{\limsup}\,\bigl| d^{\rm FOPT}_{n, \ell, p \not{=} \ell} \bigr|^{1 / n} &= \pi\,,\\
\underset{n \rightarrow \infty}{\limsup}\,\bigl| d^{\rm FOPT}_{n, \ell, p = \ell} \bigr|^{1 / n} &= \infty\,.\nonumber
\end{align}
From the $\pi^n$ global factor displayed in Eq.~\eqref{eq:dFOPTAsy} we see that the limit superior of Eq.~\eqref{eq:limsupellp} is obtained also for any linear combination of $d^{\text{FOPT}}_{n, \ell, p \not{=} \ell}$ coefficients in $\ell$ including those arising from physical weight functions where the leading $k=1$ term or potentially subleading terms for larger $k$ cancel.
Here we used that the limit superior of $|\cos(n \pi/2)|^{1/n}$ or $|\sin(n \pi/2)|^{1/n}$ can be obtained from the infinite subseries with even (or odd) $n$ for which the modulus is unity. This proves that for $\ell\neq p$ the FOPT moment series is absolutely convergent for any complex $a$ within the circle of convergence $|a|<1/\pi$.\footnote{We note that in Ref.~\cite{Gracia:2021nut} the radius of convergence of anomalous dimensions in Soft-Collinear Effective Theory and boosted Heavy-Quark Effective Theory as well as for the $\overline{\rm MS}$-mass were studied in the large-$\beta_0$ approximation. It was found that the series converge for $|a|<2.5$, which represents a much larger region of convergence than for FOPT moments with $\ell\neq p$.}
The FOPT expansion of Eq.~\eqref{eq:deltaFOPT} is therefore consistent with the OPE.
We also mention that for any region within this circle of convergence the FOPT series is also uniformly convergent according to Corollary~\ref{th:corollarypower}.
We can analytically sum the FOPT series, and for the case of monomial weight functions and positive real $a$ we find:\footnote{\label{foot:gammaasy} At first sight it may appear that the small $a$ expansion of Eqs.~\eqref{eq:FOPTsum1} and \eqref{eq:FOPTsum2} is only a divergent asymptotic series as it involves the asymptotic expansion in Eq.~\eqref{eq:gammalargez} for the incomplete gamma function for a large second argument. However, the coefficients combine to the convergent coefficients $d_{n, \ell, p\neq\ell}^{\rm FOPT}$ when all terms contributing to a single power $a^n$ are combined. The same comment also applies to the small $a$ expansions of the $H_{n,\ell}(a)$ functions in Eqs.~\eqref{eq:Hnl} and \eqref{eq:HnellfullQCDanalytic}.}
\begin{align}
\label{eq:FOPTsum1}
\delta^{(0)}_{{\rm FOPT}, 0,p\neq\ell}=& \,\frac{1}{\pi p} \arctan (a \pi) + \frac{(- 1)^p}{p} e^{- \frac{p}{a}}\! \biggl[
\frac{1}{2 \pi i} \Gamma \biggl( 0, - \frac{p}{a} - i \pi p, - \frac{p}{a} + i \pi p \biggr)\! - 1 \biggr],\\
\delta^{(0)}_{{\rm FOPT},\ell\ge 1,p\neq\ell} =&\,
\frac{1}{p - \ell} \biggl\{ e^{- \frac{\ell}{a}}\! \biggl[ \frac{1}{2
\pi i} \Gamma \biggl( 0, - \frac{\ell}{a} + i \pi \ell, - \frac{\ell}{a} - i
\pi \ell \biggr) + 1 \biggr] \nonumber \\
\label{eq:FOPTsum2}
& \,- (- 1)^{\ell + p} e^{- \frac{p}{a}}\! \biggl[ \frac{1}{2 \pi i} \Gamma
\biggl( 0, - \frac{p}{a} + i \pi p, - \frac{p}{a} - i \pi p \biggr) + 1 \biggr] \!\biggr\}\,.
\end{align}
For illustration, we have displayed the truncated moment series $\delta^{(0)}_{{\rm FOPT},\ell,p\neq\ell}$ for the values $(\ell,p)=(0,2),(1,2)$
in Fig.~\ref{fig:FOPTvsCIPT} as red dots using $a=0.2256$ which corresponds to \mbox{$\alpha_s^{(n_f=3)}(s_0)=0.315$}. The dashed horizontal lines indicate the series sums.
The visible oscillations of the FOPT series are substantially damped for physical weight functions due to the cancellation of the leading asymptotic contributions mentioned before.

We note that since the size of the strong coupling $\alpha_s$ is only known with an uncertainty and furthermore depends on the observable (through the value of $s_0$), we do not discuss the particular case $|a|=1/\pi$. This case does not have any specific meaning in practice. So when we talk about convergent series, we always refer to absolutely convergent series in the rest of this article, and it is the fact that a finite interval of convergence for $a$ exists that matters for the consistency with the OPE. We also acknowledge that an indirect proof of the convergence using the Borel representation of the spectral function moment series based on Eq.~\eqref{eq:invBorelD} with the unexpanded Borel function $B_p(u)=1/(p-u)$ for $p=2,3,\ldots$ and using the renormalon calculus has been given in the Appendix of Ref.~\cite{Benitez-Rathgeb:2022yqb}. The proof presented above directly deals with the actual series.\footnote{The proof based on the Borel transform in Ref.~\cite{Benitez-Rathgeb:2022yqb} also applies in full QCD.}

\subsection{CIPT Series}
\label{sec:W1CIPT}

In the large-$\beta_0$ approximation the CIPT expansion functions $H_{n, \ell}(a)$ can be readily computed by rewriting the integral of Eq.~\eqref{Hndef} in terms of the phase angle $x=e^{i\phi}$, which gives
\begin{equation}
\label{Hndefb0}
H_{n, \ell}(a) = \frac{1}{2\pi}\!\int_{-\pi}^{+\pi}
\!\! \dd \phi \, e^{i\ell\phi} \biggl(\frac{a}{1+i a \phi}\biggr)^{\!\!n}
= \frac{i}{\pi} e^{-\frac{\ell}{a}}\!\!\int_{t_-}^{t_+} \!\! \dd t \, (-2t)^{-n}e^{-2\ell t}\,,
\end{equation}
where the integral in the second equality is obtained by a change of variable to $t=-\frac{1}{2a(-x s_0)}$ with the integration boundaries $t_\pm=-\frac{1}{2a(-s_0\pm i0)}=-\frac{1}{2a_\pm}=-\frac{1\pm i a \pi}{2a}$. The $H_{n,\ell}(a)$ functions have poles at $a=\pm i/\pi$ since the phase integral becomes singular at one of the boundaries $\pm\pi$ (or one of the $t_\pm$ is zero and starts at the pole singularity at $t=0$).
We define the phase integral in the interval $[-\pi,+\pi]$ strictly along the real axis. Since $\mbox{Re}[t_\pm]=-\frac{\mbox{\footnotesize Re}[a]}{2|a|^2}$ and
$\mbox{Im}[t_\pm]=\frac{1}{2}\bigl(\frac{\mbox{\footnotesize Im}[a]}{|a|^2}\mp \pi\bigr)$ this corresponds to a complex path in $t$ that is in the straight vertical downward direction.
The definition with a fixed $\phi$ integration path, where the distance to the origin at $\phi=0$ is bounded, ensures that $H_{n, \ell}(a)$ is analytic at least in some finite neighborhood of the origin $a=0$ and that
$\lim_{a\to 0} H_{n, \ell}(a)=0$ from all directions in the complex $a$ plane. In fact, at the origin the $H_{n,0}(a)$ vanish like $a^n$, while the $H_{n,\ell\ge 1}(a)$ vanish like $a^{n+1}$ because of the complex phase $e^{i\ell\phi}$ in Eq.~\eqref{Hndefb0}. In other words, $H_{n,\ell}(a)={\cal O}(a^{n+1-\delta_{\ell,0}})$ as $a\to 0$.\footnote{\label{foot:a}This is associated to the fact that the ${\cal O}(a)$ FOPT term in the moments $\delta_{{\FOPT}, \ell\ge 1}^{(0)}$ always vanish. The absence of this term for $\ell\ge 1$ entails some notation subtleties in expressions we present later in this article.}

Our path prescription provides an unambiguous definition of the integrals for all complex $a$ besides where there are cuts. Such cuts arise when the strong coupling Landau pole at $\phi=i/a$ traverses through the path of $\phi$ that goes along the real axis from $-\pi$ to $+\pi$ (or the straight path of $t$ from $t_-$ to $t_+$ traverses through $t=0$). With our definition these cuts are located along the straight lines $(-i\infty,-i/\pi)$ and $(+i/\pi,+i\infty)$ on the imaginary axis of the complex $a$ plane. Any other path definition would lead to the same poles and branch points at $a=\pm i/\pi$, but to a different curve of the cut connecting the branch points. A path for $\phi$ deformed into the positive (negative) imaginary plane corresponds to a path for $x$ with $|x|\le 1$ ($|x|\ge 1$) in Eq.~\eqref{Hndef}. Thus, if the $\phi$ integration contour would be deformed far away from the real axis, it is possible that the minimal distance of the cut curve to the origin at $a=0$ becomes smaller than $1/\pi$. Our definition is the most obvious one, as it is the standard path used for real $a$, see Eq.~\eqref{Hndef}, and leads to branch cuts along the imaginary axis away from the origin. However, these cuts only truly arise when the Landau pole leads to a simple pole in the integrands in Eq.~\eqref{Hndefb0}, i.e., when there is a finite residue that can make a contribution when the Landau pole at $\phi=i/a$ crosses the integration path. Since the residue has the general form $\frac{1}{2 \pi i}\frac{\ell^{n-1}}{\Gamma(n)}e^{-\ell/a}$, cuts only arise for $(n,\ell)=(1,0)$ or when $\ell\ge 1$. There are no cuts for $\ell=0$ and $n\ge 2$. Except for the poles and the cuts (if they arise), the $H_{n,\ell}(a)$ are analytic in the entire complex $a$ plane. In any case, for any sensible contour path choice (with $\phi$ being close to the real axis or $|x|$ being equal or close to $1$) in Eq.~\eqref{Hndef}] the $H_{n,\ell}(a)$ functions are all analytic within the circle with radius $1/\pi$ around the origin, where they can also be expanded in a convergent $a^n$ Taylor series.

The analytic results for complex $a$ read
\begin{align}
\label{eq:H10}
H_{1,0}(a) =\, & \frac{1}{2 \pi i} \log \biggl( \frac{a_-}{a_+} \biggr)\!
= \frac{1}{\pi}\arctan(\pi a) \, ,\\
H_{n\geq2,0}(a) =\, & \frac{a_-^{n - 1} - a_+^{n-1}}{2 \pi i (n -1)} =
\label{eq:Hn0}
\biggl(\frac{a}{\sqrt{1+a^2\pi^2}}\biggr)^{\!\!n-1}\frac{\sin[(n-1)\arctan(a\pi)]}{\pi(n-1)} \,, \\
H_{n, \ell\ge 1} (a) =\,&
\frac{(- 1)^{\ell}}{\pi} \sum_{k = 1}^{n - 1} \frac{\ell^{k-1}}{(n-k)_k} \biggl(\frac{a}{\sqrt{1+a^2\pi^2}}\biggr)^{\!\!n-k} \sin [(n-k) \arctan (\pi a)] \nonumber\\
& + \frac{\ell^{n-1} e^{-
\frac{\ell}{a}}}{\Gamma(n)}\! \biggl[ \frac{1}{2 i \pi} \Gamma \biggl( 0, -\frac{\ell}{a_+},-\frac{\ell}{a_-}\biggr) +\Theta(a) \biggr]\nonumber\\
=\,&
\frac{(- 1)^{\ell}}{\pi} \sum_{k = 1}^{n - 1} \frac{\ell^{k-1}}{(n-k)_k} \frac{a_-^{n - k} - a_+^{n-k}}{2 \pi i }
+ \frac{\ell^{n-1} e^{-
\frac{\ell}{a}}}{\Gamma(n)}\! \biggl[ \frac{1}{2 i \pi} \Gamma \biggl( 0, -\frac{\ell}{a_+},-\frac{\ell}{a_-}\biggr) +\Theta(a) \biggr]\nonumber \\
\label{eq:Hnl}
= \,& e^{-\frac{\ell}{a}} \ell^{n-1}\!\biggl[\frac{(-1)^n}{2i\pi}\Gamma\biggl(1-n,-\frac{\ell}{a_+},-\frac{\ell}{a_-}\biggr)\!+
\frac{\Theta(a)}{\Gamma(n)}\biggr]\,,
\end{align}
with
\begin{equation}
\label{eq:Theta}
\Theta(a)=\theta [{\rm Re} (a)]\theta \biggl( \biggl| a + \frac{i}{2 \pi} \biggr| -
\frac{1}{2 \pi} \biggr) \biggl[ 1- \theta \biggl( \frac{1}{2 \pi} - \biggl| a - \frac{i}{2 \pi} \biggr|
\biggr) \!\biggr]\,,
\end{equation}
where in addition we have $H_{n, \ell} (0)=0$ since one cannot simply insert $a=0$ in Eq.~(\ref{eq:Hnl}). The term proportional to $\Theta(a)$ arises from the definition of the incomplete gamma function in Eq.~\eqref{eq:incGammaDef} which otherwise leads to a mismatch related to the residue at $t=0$ already quoted above. We remind the reader that the residue becomes relevant due to the use the incomplete gamma functions, which implies a path deformation in the $t$ variable through real $+\infty$. The step function $\theta(x)$ appearing in
$\Theta(a)$ is defined as $\theta(x)=1$ for $x\geq 0$ and $\theta(x)=0$ for $x<0$. We note that the particular dependence of $\Theta(a)$ on the step function $\theta(x)$-definition at $x=0$ arises due to the convention used for the incomplete gamma function when the second argument is negative real. The result as shown above arises due to the definition $\Gamma(a,x)\equiv \Gamma(a,x+i0)$ for negative real $x$, as it is used e.g.\ in Wolfram Mathematica~\cite{Mathematica}.
We note that the analytic evaluation of the integrals in Eq.~(\ref{Hndefb0}) in terms of the incomplete gamma function (where the $t$ path is always deformed through positive real infinity) implies that for values of $a$ on the cut curves, i.e.\ when $i/a$ is in the interval $(-\pi,+\pi)$, the $\phi$ integration is deformed around the pole into the positive imaginary half-plane.
Also note that for the first two equalities in Eq.~\eqref{eq:Hnl} we have used Eq.~\eqref{eq:recursive}. Interestingly, they provide the asymptotic expansion of $H_{n, \ell\ge 1}(a)$ as $n\to\infty$, where the $k=1$ term represents the leading contribution.

\begin{figure*}[t]
\subfigure[]{
\includegraphics[width=0.44\linewidth]{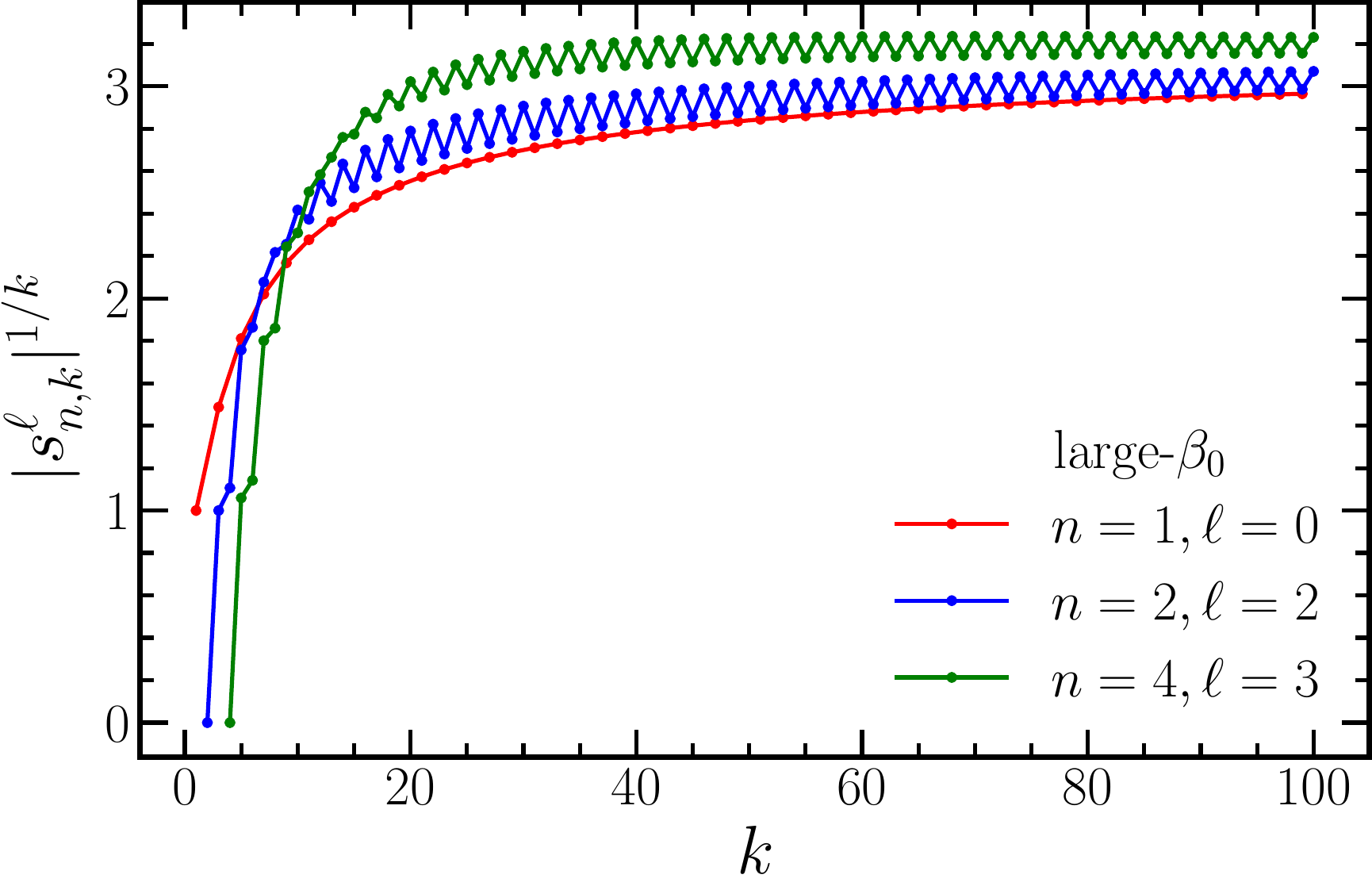}
\label{fig:Largeb0S}}
\subfigure[]{\includegraphics[width=0.45\linewidth]{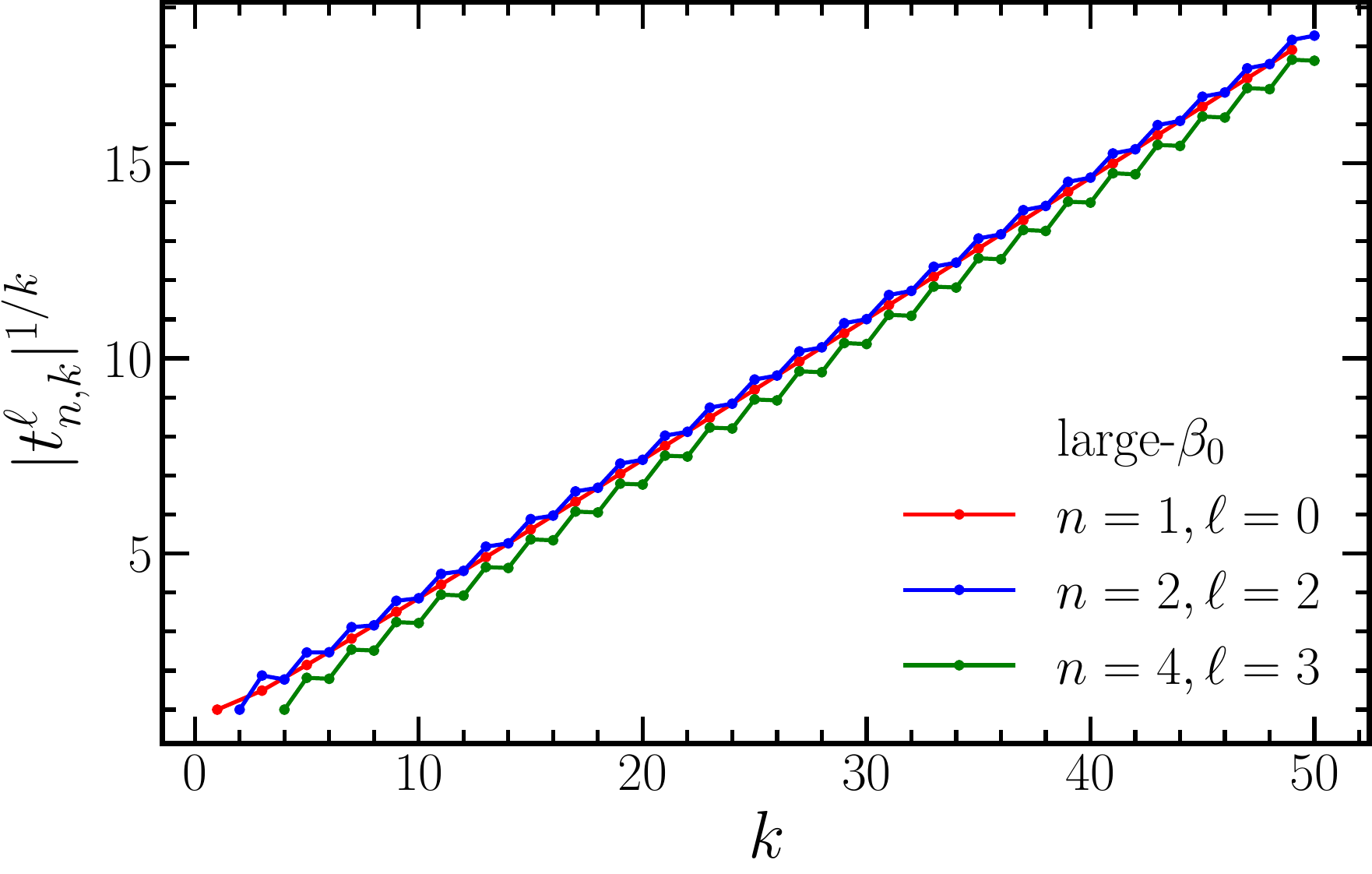}
\label{fig:Largeb0T}}
\caption{The $k$-th root of the
$s$ and $t$ expansion coefficients $|s^\ell_{n,k}|^{1/k}$ (panel a) and $|t^\ell_{n,k}|^{1/k}$ (panel~b) as a function of $k$, to visualize the Cauchy root test concerning the convergence of the series in Eqs.~\eqref{eq:Hninak} and \eqref{eq:aninHk}. Red, blue, and green refer to $(n,\ell)=(1,0)$, $(2,2)$, and $(4,3)$, respectively.}
\label{fig:ST}
\end{figure*}

The analytic properties of the functions $H_{n, \ell}(a)$ imply that they can be expanded in Taylor series around $a=0$ which are absolutely convergent for $|a|<1/\pi$:
\begin{equation}
\label{eq:Hninak}
H_{n, \ell}(a) = \sum_{k=n}^\infty s_{n,k}^{\ell} a^k\,,
\end{equation}
The expansion coefficients $s_{n,k}^{\ell}$ are directly related to the expressions for $I_{k, \ell}$ defined in Eq.~\eqref{eq:Ikdef} and read
\begin{equation}
\label{eq:snkell}
s^\ell_{n,k}
= \frac{(- 1)^{k + n} I_{k - n,\ell}}{\Gamma (n) (k)_{1 - n}} \,,
\end{equation}
where we note that $s_{n, n}^{\ell}=\delta_{\ell,0}$. Therefore, the $s_{n, k}^{\ell}$ form upper-triangular matrices if we take $n$ and $k$ as the row and column indices, respectively.
Using the third equality of Eq.~\eqref{eq:InlIntegrals}, we also find the infinite series representation
\begin{equation}
\label{eq:snkasy}
s_{n, k}^{\ell} =
\frac{ (- 1)^{\ell} \pi^{k - n}}{\Gamma (n)}
\sum_{j = 0}^{\infty} \frac{(\ell \pi)^j}{(k)_{j-n+2}}\cos \biggl[ \frac{\pi}{2} (j + k - n)
\biggr]\,,
\end{equation}
which is absolutely convergent, but also provides an (asymptotic) expansion as $k\to\infty$, where the $j=0$ term is the leading contribution.
For $\ell=0$ only the first term in the $j$-sum survives.
Since there is an infinite subsequence in $k$ such that the modulus of the cosine functions is in the interval $[\delta,1]$ for a $0<\delta<1$ and using Stirling's formula for the gamma functions (contained in the Pochhammer symbol) we find
\begin{equation}
\limsup_{k\to\infty} |s^\ell_{n,k}|^{1/k}=\pi\,.
\end{equation}
This reconfirms our statement above concerning the convergence radius of the Taylor expansion around the origin of the $H_{n, \ell}$ functions in the complex $a$ plane.
For illustration we have displayed the results for $|s^\ell_{n,k}|^{1/k}$ as a function of $k$ in Fig.~\ref{fig:Largeb0S} for various combinations of $(n,\ell)$.
We have explicitly checked that summing the series in Eq.~\eqref{eq:Hninak} indeed converges to the analytic expression in Eqs.~\eqref{eq:H10}, \eqref{eq:Hn0}, and \eqref{eq:Hnl} for all complex $a$ with $|a|<1/\pi$.
It is an interesting fact that, similar to the FOPT coefficients $d^{\text{FOPT}}_{n, \ell, p \not{=} \ell}$, as discussed following Eq.~\eqref{eq:dFOPTAsy}, the leading
and potentially subleading asymptotic terms cancel in linear combinations of $\ell$ that arise for physical weight functions $W(x)$, but the limit superior remains $\pi$ even for such linear combinations due to the global factor of $\pi^{k}$ shown in Eq.~\eqref{eq:snkasy}.

From the expressions in Eqs.~\eqref{eq:Hn0} and \eqref{eq:Hnl} it is straightforward to determine the limit superior of $|H_{n,\ell}(a)|^{1/n}$ as $n\to\infty$. The result reads
\begin{equation}
\label{eq:Hlimsup}
\limsup_{n\to\infty} |H_{n,\ell}(a)|^{1/n} = \max(|a_+|,|a_-|) =\frac{|a|}{\sqrt{1+|a|^2\pi^2-2\pi |\mathrm{Im}(a)|}}\,.
\end{equation}
Interestingly, we yet again find that the leading (and potentially subleading) asymptotic contributions as $n\to\infty$ cancel in linear combinations of the $H_{n,\ell}$ functions that arise for physical weight functions $W(x)$ in analogy to the discussion following Eqs.~\eqref{eq:dFOPTAsy} and ~\eqref{eq:snkasy}. The limit superior given in Eq.~\eqref{eq:Hlimsup} also applies for these linear combinations, as we can see from the form of the subleading asymptotic terms given in Eq.~\eqref{eq:Hnl}.

\begin{figure}[t]
\centering
\includegraphics[width=0.7\linewidth]{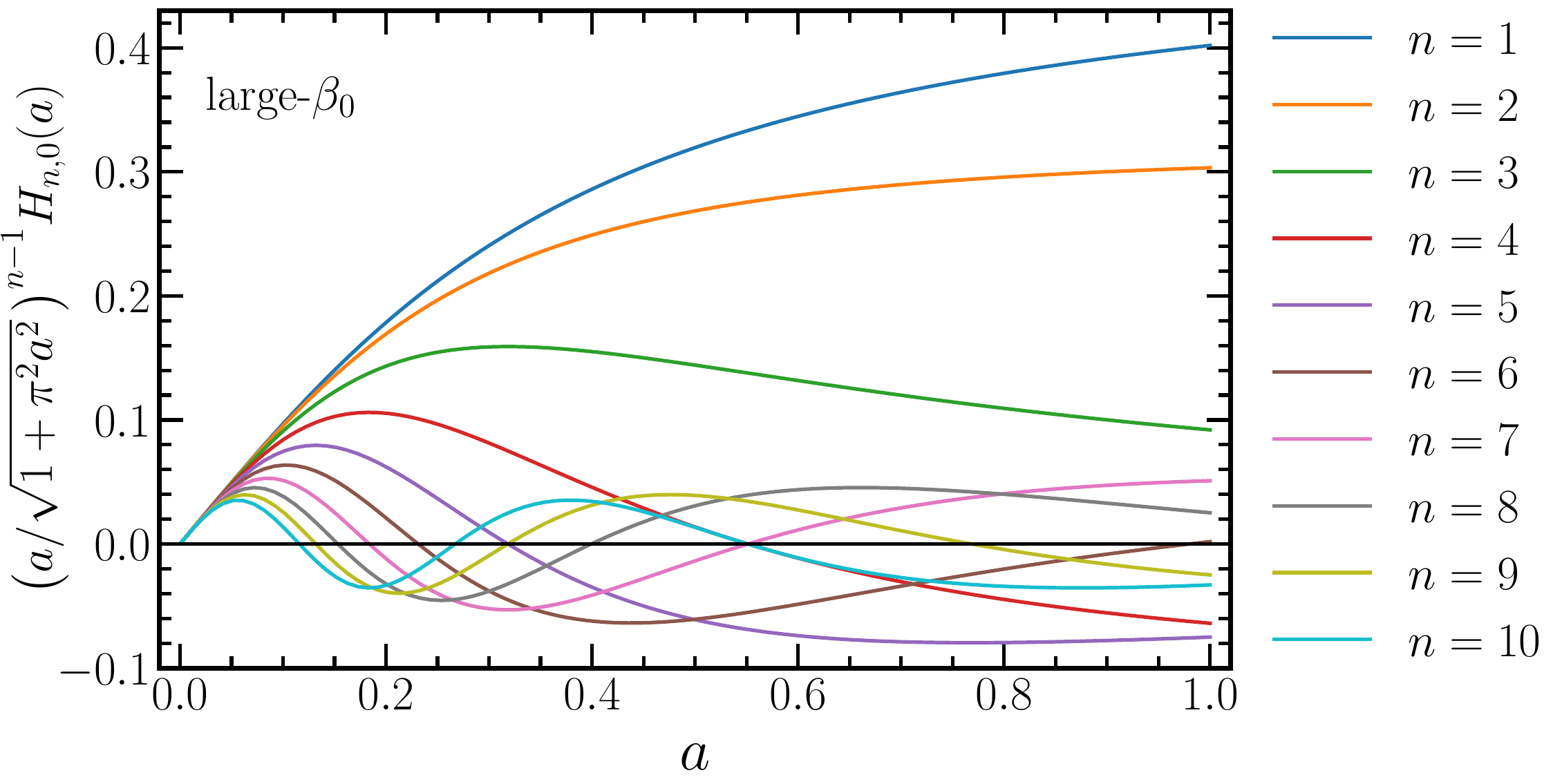}
\caption{The functions $H_{n,0}(a)/\bigl(a/\sqrt{1+a^2\pi^2}\,\bigr)^{\!n-1}$ for $1\leq n\leq 10$ illustrating that they are bounded in the interval $[0,1]$.}
\label{fig:HBoundaryLargeb0}
\end{figure}

From Eqs.~\eqref{eq:Hn0} and \eqref{eq:Hlimsup} we see that the $n$ scaling of the $H_{n,\ell}$ functions is in powers of $a/\sqrt{1+a^2\pi^2}$. The fact that $|a|/\sqrt{1+a^2\pi^2}<\min(|a|,1/\pi)$ for real $a$, is one of the reasons why the CIPT expansion typically exhibits a better behavior of the series at low orders than the power $a^n$ expansion of FOPT, see also Fig.~\ref{fig:HBoundaryLargeb0}, were $H_{n,0}(a)/(a/\sqrt{1+a^2\pi^2})^{n-1}$ is shown for $1\leq n\leq 10$ in the interval $[0,1]$.
It is clearly visible that this scaling property does not only arise for large $n$, but is active for all $n$ values.

It is now straightforward to examine the convergence of the single pole $1/(p-u)$ renormalon series contribution in the Adler function to the CIPT moment series $\delta^{(0)}_{{\CIPT},\ell,p\neq\ell}$ in Eq.~\eqref{eq:deltaCIPTdef}. Using the leading asymptotic expressions of the coefficients $\bar{c}_{n,1,p}=\Gamma(n)/p^{n}$ from Eq.~(\ref{eq:cnkb0}) and the result of Eq.~\eqref{eq:Hlimsup}, we find
\begin{equation}
\label{eq:CIPTlimsup}
\limsup_{n\to\infty} |\bar{c}_{n,1,p} H_{n,\ell}(a)|^{1/n}=\infty\,.
\end{equation}
We see that the limit superior diverges for any value of $a$. The outcome is the same for any linear combination of $H_{n,\ell}$ functions in $\ell$ including those arising from the physical weight functions as can be seen from the form of the subleading asymptotic terms in Eq.~\eqref{eq:Hnl}.

We have displayed the truncated moment series $\delta^{(0)}_{{\CIPT},\ell,p\neq\ell}$, in Fig.~\ref{fig:FOPTvsCIPT} for $(\ell,p)=(0,2),(1,2)$ and $a=0.2256$ as the blue dots.
We can clearly see that CIPT shows a much better apparent behavior at low orders than FOPT (red dots). The CIPT series does in particular not show the oscillatory behavior of FOPT.
This is due to the scaling of the $H_{n,\ell}$ functions just mentioned above, as well as the zeros of the $H_{n,\ell}$ functions visible in Fig.~\ref{fig:HBoundaryLargeb0}, which further suppress the size of $H_{n,\ell}(a)$. These zeros, which appear to have a considerable benefit at this point, turn out to play an important additional role in the further considerations of this article.
However, the fact that the CIPT series is divergent for any value of $a$ renders $\delta^{(0)}_{{\CIPT},\ell,p\neq\ell}$ inconsistent with the OPE. The consistency with the OPE demands that the series must be convergent for $\ell\neq p$. The visible discrepancy\footnote{This discrepancy has been called the asymptotic separation and has been quantified analytically in Refs.~\cite{Hoang:2020mkw,Hoang:2021nlz}.} between the value the CIPT series appears to approach at intermediate orders $10 < n < 15$ and the (correct physical) value of the FOPT series (red dashed horizontal line) renders the CIPT method phenomenologically inconsistent because the Adler function series contains IR renormalon contributions.

We want to remind the reader, however, that CIPT can still be used as a viable phenomenological method once the dominant IR renormalons (with the smallest values of $p$) contained in the Adler function are removed by subtractions~\cite{Benitez-Rathgeb:2022yqb,Benitez-Rathgeb:2022hfj} as the problematic aspects of CIPT just discussed can then be suppressed to a negligible level in the presence of the experimental uncertainties in $\tau$-decay spectral data and truncations of the perturbative series. In the absence of any factorially divergent behavior in the coefficients $\bar c_{n,1}$, such that $\limsup_{n\to\infty}|\bar{c}_n|$ would be finite (which is not true for the Adler function), the CIPT series in Eq.~\eqref{eq:deltaCIPTdef} would have a finite interval of convergence and also yield not only an absolutely but also a uniformly convergent series in any subinterval of the interval of convergence. The important property of uniform convergence in such a situation\footnote{We note that the uniformity of the convergence of a series in functions $\phi_1(x), \phi_2(x), \ldots$ and the property of uniformity of the asymptotic sequences $\{\phi_n(x)\}$ discussed below are two different mathematical aspects that must not be confused.} can be seen from Eqs.~\eqref{eq:Hn0} and \eqref{eq:Hnl}. The expressions show that for real $a$ the functions $H_{n,\ell}(a)$ [or any physical linear combination of $H_{n,\ell}(a)$] are bounded by $\bigl(|a|/\sqrt{1+a^2\pi^2}\,\bigr)^{\!n-1}$ for $n$ larger than some integer $\bar n$. This is illustrated in Fig.~\ref{fig:HBoundaryLargeb0} for the functions $H_{n,0}(a)$, which are bounded by the function $\bigl(|a|/\sqrt{1+a^2\pi^2}\,\bigr)^{\!n-1}$ for all $n\in\mathbb{N}$. Since the latter expression is bounded in any subinterval, that CIPT series would also be uniformly convergent in any such subinterval according to the Weierstrass criterion for uniform convergence in Theorem~\ref{th:Weierstrassuniform}.

\begin{figure*}[t!]
\centering
\subfigure[]{\includegraphics[width=.404\linewidth]{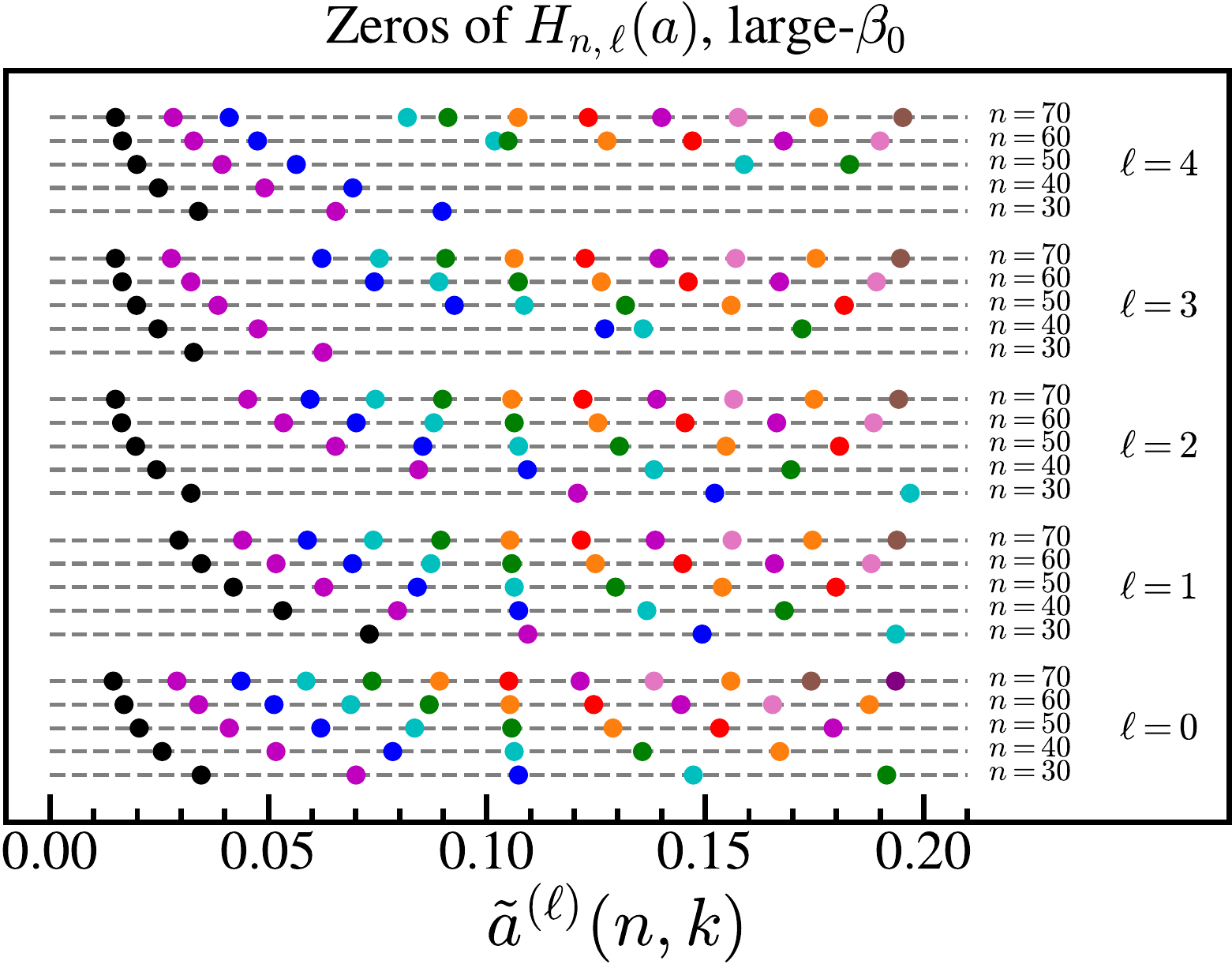}
\label{fig:ZerosHnl1}}
\subfigure[]{\includegraphics[width=.4648\linewidth]{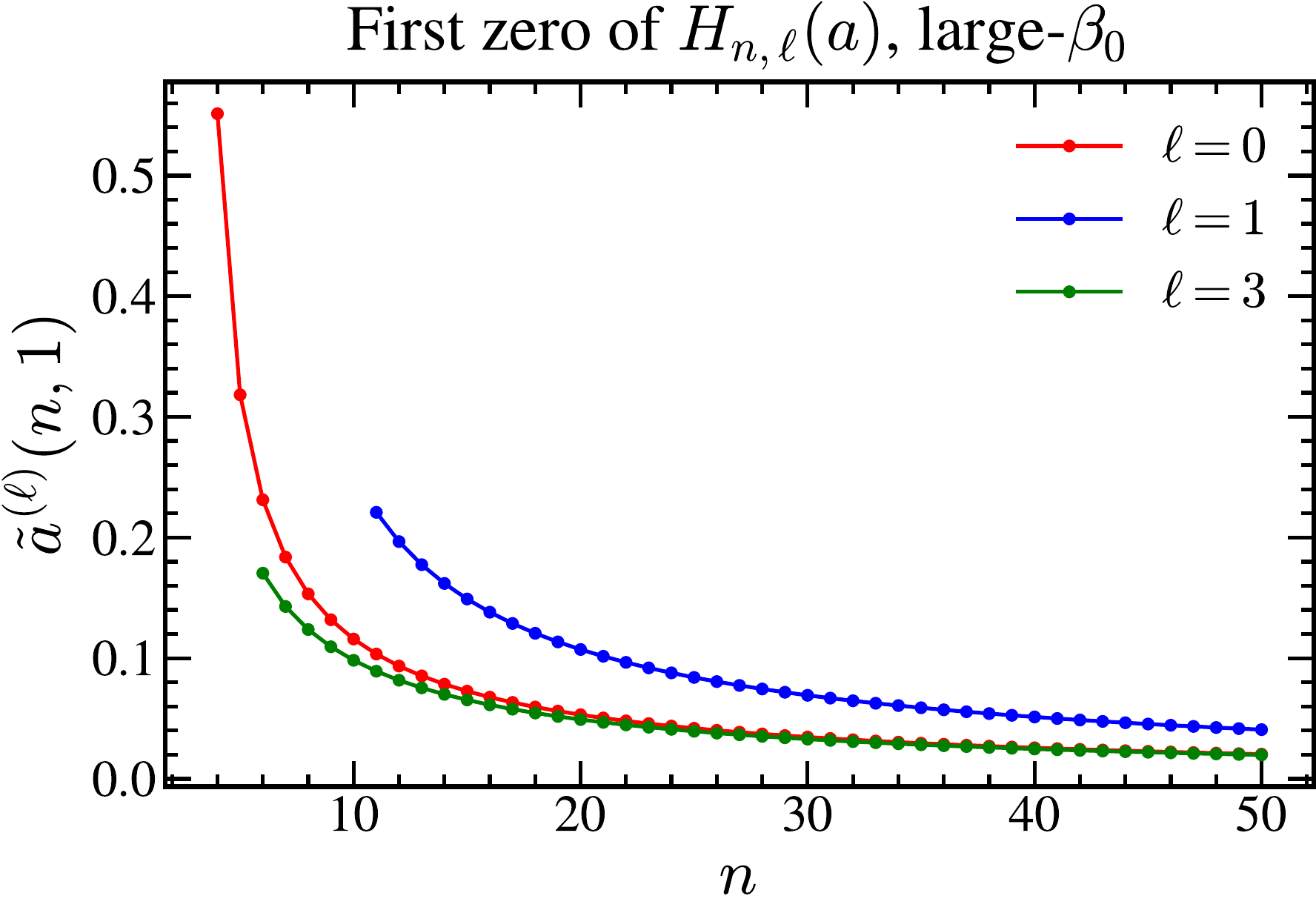}
\label{fig:ZerosHnl2}}
\caption{Positive zeros of $H_{n,\ell}(a)$. All zeros are located on the real axis. Panel a: Position of the zeros of $H_{n,\ell}(a)$ in the interval $(0,0.2)$. Each horizontal line corresponds to a value of $n$, and lines are grouped according to $\ell$.
Panel b: Position of the positive zero closest to the origin as a function of $n$ for $\ell=0$ (blue),
$\ell=1$ (red), and $\ell=3$ (green).}
\label{fig:ZerosHnl}
\end{figure*}

We conclude this section with the confirmation that the CIPT expansion in the functions $H_{n, \ell}(a)$ in Eq.~\eqref{eq:deltaCIPTdef} is indeed an asymptotic expansion from the mathematical perspective. Even though the CIPT expansion has now been used in phenomenological analyses for a number of decades, it is useful to proceed with the discussion on the mathematical aspects of CIPT at this basic level. Since the following considerations are somewhat mathematical, we refer the reader not familiar with the quoted definitions and theorems to Appendix~\ref{sec:asymptoticheorems}.

According to the general definition of an asymptotic expansion as $a\to 0$, quoted in Definition~\ref{th:asymptexpansion}, it is mandatory that the $H_{n, \ell}$ functions form an asymptotic sequence $\{H_{n, \ell}(a)\}=\{H_{1, \ell}(a),H_{2, \ell}(a),\ldots\}$ as $a\to 0$
for any complex domain $R$ containing the origin as a limit point. This means that they
obey the ``little $o$'' order relation $H_{n+1, \ell}=o(H_{n, \ell})$ as $a\to 0$ for all $n\in\mathbb{N}$ and for any $\ell\in\mathbb{N}_0$, see Definition~\ref{th:asymptsequence}. The definition of the little-order relation is quoted in Definition~\ref{th:littleo}. The property of an asymptotic sequence then ensures that the coefficients of the asymptotic expansion can be determined in an unambiguous way by the recursion relation formulated in Corollary~\ref{th:determinean}. Since the $H_{n, \ell}$ functions are nonzero in at least some neighborhood of $a=0$ (excluding $a=0$, where the $H_{n, \ell}$ functions vanish), the little-order relation already follows from the property
\begin{equation}
\label{eq:elHratio}
\lim_{a\to 0}\frac{|H_{n+1,\ell} (a)|}{|H_{n,\ell}(a)|}=0\,,
\end{equation}
for any $n\in\mathbb{N}$. This can be seen from the form of their Taylor expansions in Eq.~(\ref{eq:Hninak}) and the fact that the $H_{n,0} (a)$ vanish like $a^n$ and the $H_{n,\ell\ge 1} (a)$ vanish like $a^{n+1}$ as $a\to 0$.

However, the $\{H_{n,\ell}(a)\}$ form {\it nonuniform} asymptotic sequences, see Definition~\ref{th:uniasymptsequence}, as one has to consider smaller and smaller domains around the origin $a=0$ such that the ratio shown in Eq.~\eqref{eq:elHratio} is small. This is due to the zeros already mentioned above and also visible in Fig.~\ref{fig:HBoundaryLargeb0}. To see this at the analytic level let us first consider the case $\ell = 0$.
The ratio of two consecutive $H_{n,0}$ functions reads
\begin{equation}
\label{eq:Hnratio}
\frac{H_{n+1,0}(a)}{H_{n,0}(a)}=\frac{a(n-1)}{n\sqrt{1+a^2\pi^2}}\frac{\sin[n\arctan(a\pi)]}{\sin[(n-1)\arctan(a\pi)]}\,.
\end{equation}
From the oscillatory properties of the sine function
it is easy to see that, if $R$ contains intervals of the real axis (containing the origin or having the origin as a limit point), there is no neighborhood $U_\epsilon$ of $a=0$ such that the ratio in Eq.~\eqref{eq:elHratio} is smaller than a given small $\epsilon$ for all $n$. In other words, there always exists such a neighborhood $U^{(n)}_\epsilon$ for any $n$, but the maximal distance of the points in $U^{(n)}_\epsilon$ to $a=0$ shrinks to zero as $n$ increases if $a$ can be real.
This is because the $\{H_{n,0}(a)\}$ have zeros besides $a=0$ along the real $a$ axis, which happen to approach zero as $n$ increases. These zeros arise from the
phase oscillations due to the $i a \phi$ term in Eq.~\eqref{Hndefb0} and follow the condition $(n-1)\arctan(a\pi)=k\pi$ with $k\in\mathbb{Z}$ which holds only if $\arctan(a\pi)\in\mathbb{R}$. Concretely, they are located at
$\tilde{a}^{(0)}(n,k)=\frac{1}{\pi}\tan \bigl(\frac{k\pi}{n-1}\bigr)$ for $-\frac{n-1}{2}<k<\frac{n-1}{2}$, where $k=0$ corresponds to $H_{n, \ell} (0)=0$. While the number of zeros increases with $n$, the zeros $\tilde{a}^{(0)}(n,k\neq 0)$ all vanish like $1/n$ as $n\to\infty$ for any $k$.
The zeros are displayed graphically in Fig.~\ref{fig:ZerosHnl} exemplarily for a number of $n$ values.
For $\ell>0$ zeros with analogous qualitative characteristics $\tilde{a}^{(\ell)}(n,k)$ arise as well since the factor $(-x)^\ell$ in Eq.~\eqref{Hndefb0} only provides an additional modulation of the phase cancellations. However, finding an analytic expression for the zeros is harder. We have displayed them in Fig.~\ref{fig:ZerosHnl} as well for a number of $\ell$ values obtained from numerical evaluations. An approximate
analytic insight on the zeros for $\ell>0$ can, however, be gained with the large-$n$ asymptotic behavior of the $H_{n,\ell\ge 0}(a)$ functions as $n\to\infty$ already given in Eq.~\eqref{eq:Hnl}.

\section{Having a deeper Look}
\label{sec:W1Deeper}

After having discussed the analytic properties of the CIPT expansion functions $H_{n,\ell}(a)$ in the previous section we now proceed to have a deeper look into the mathematical circumstances why the expansion in terms of the asymptotic sequences $\{H_{n,\ell}(a)\}$ is divergent in cases where physics requires a convergent perturbative series. The aim is to identify the mathematical property of the functions $H_{n,\ell}(a)$ that leads to this behavior, independently of the application in the context of the $\tau$ hadronic spectral function moments. We remind the reader that the motivation of this analysis is to identify mathematical criteria to scrutinize also other expansion methods where integrations over the renormalization scale are employed.

To this end we consider transformations of the CIPT or FOPT series by reexpanding one of the series in terms of the
other expansion functions. We thus also need the expansion of any power $a^n$ in terms of the $H_{k, \ell}(a)$ functions for each $\ell=0,1,2,\ldots$ in addition to the reverse expansion already given in Eq.~\eqref{eq:Hninak}. It has the form
\begin{equation}
\label{eq:aninHk}
a^n = \sum_{k=n+(\delta_{\ell,0}-1)}^\infty t_{n,k}^{\ell} H_{k, \ell}(a)\,.
\end{equation}
Note that for $\ell\ge 1$ the relation only matters for $n=2,3,\ldots$, and the Kronecker delta $\delta_{\ell,0}$ shown here and below arises for the reason explained in Footnote~\ref{foot:a} in order to make terms that do not contribute explicit.

So we consider the series transformations
\begin{align}
\label{eq:CIPTtransform}
\sum^\infty_{n = 1} c_n H_{n, \ell}(a)
& \longrightarrow
\sum^\infty_{n = 1}\sum^\infty_{k = n+\delta_{\ell,0}}\!\! c_n s_{n,k}^{\ell} a^k \,,
\\
\label{eq:FOPTtransform}
\sum^\infty_{n = 2-\delta_{\ell,0}}\!\! d_n a^n
& \longrightarrow
\sum^\infty_{n = 2-\delta_{\ell,0}} \sum^\infty_{k = n+(\delta_{\ell,0}-1)}\!\! d_n t_{n,k}^{\ell} H_{k, \ell}(a) \,,
\end{align}
which represent the transformation of a single series into a double series. Note that at this point our notation does not imply any particular ordering prescription for the summation of the double series. So one should think of the emerging double series as two-dimensional arrays. For example, we have
\begin{equation}
\label{eq:arrayformal}
\sum^\infty_{n = 1} \sum^\infty_{k = n} e_{n,k} \; \sim \;
\left(
\begin{array}{cccc}
e_{1,1} & e_{1,2} & e_{1,2} & \cdots \\
0 & e_{2,2} & e_{2,3} & \cdots \\
0 & 0 & e_{3,3} & \cdots \\
\vdots & \vdots & \vdots & \ddots
\end{array}\right),
\end{equation}
where the horizontal lines arise from the series of the expansion function used for the series prior to the double series transformation. A particular summation prescription is indicated explicitly by additional parentheses. Thus, in the example $\sum^\infty_{n = 1}(\sum^\infty_{k = n} e_{n,k})$ stands for summing the horizontal series first and the results of these horizontal sums afterwards. On the other hand
$\sum^\infty_{k = 1} (\sum^k_{n = 1} e_{n,k})$ stands for summing the vertical series first and the results of these vertical sums afterwards. For the cases discussed below, the limits of the sums are a bit more involved due to the appearance of $\delta_{\ell,0}$, which is, however, not significant for the main purpose of the discussion.

\subsection{Double Series Transformations}
\label{sec:doublesum}
We start by discussing the transformation of a CIPT series into a double series of power terms, see Eq.~\eqref{eq:CIPTtransform}.
Consider a CIPT series $\sum^\infty_{n = 1}c_n H_{n, \ell}(a)$ that is convergent according to the root criterion for some $a=a_0$ within the region of convergence. It is then also absolutely convergent as well as uniformly convergent at least in the circle around zero with radius $|a_0|$. The property of uniform convergence follows from the fact that the $H_{n, \ell}(a)$ are bounded as we discussed already at the end or Sec.~\ref{sec:W1CIPT}.
In addition, the horizontal lines from the expansion of the $H_{n, \ell}(a)$ functions in powers of $a$ are absolutely convergent in the circle around zero with radius $r<1/\pi$. According to the Weierstrass double series Theorem~\ref{th:Weierstrassdouble} these conditions are sufficient so that doing either horizontal or vertical sums first leads to the same result,
$\sum^\infty_{n = 1} (\sum^\infty_{k = n+\delta_{\ell,0}} c_n s_{n,k}^{\ell} a^k) =
\sum^\infty_{k = 1+\delta_{\ell,0}}(\sum^{k-\delta_{\ell,0}}_{n = 1} c_n s_{n,k}^{\ell} a^k)$ for $|a| < \mbox{max}(a_0,r)$. In other words, an absolutely convergent CIPT series can be reexpanded into a convergent FOPT series which sums to the same value. Furthermore, from the discussions in Sec.~\ref{sec:W1largeb0} we also know that it is possible that a divergent CIPT series can be reexpanded into an absolutely convergent FOPT series. So starting from a CIPT series and reexpanding it in FOPT does not make its convergence property worse, but can even improve it.

\begin{figure*}[tb!]
\subfigure[]{\includegraphics[width=0.445\linewidth]{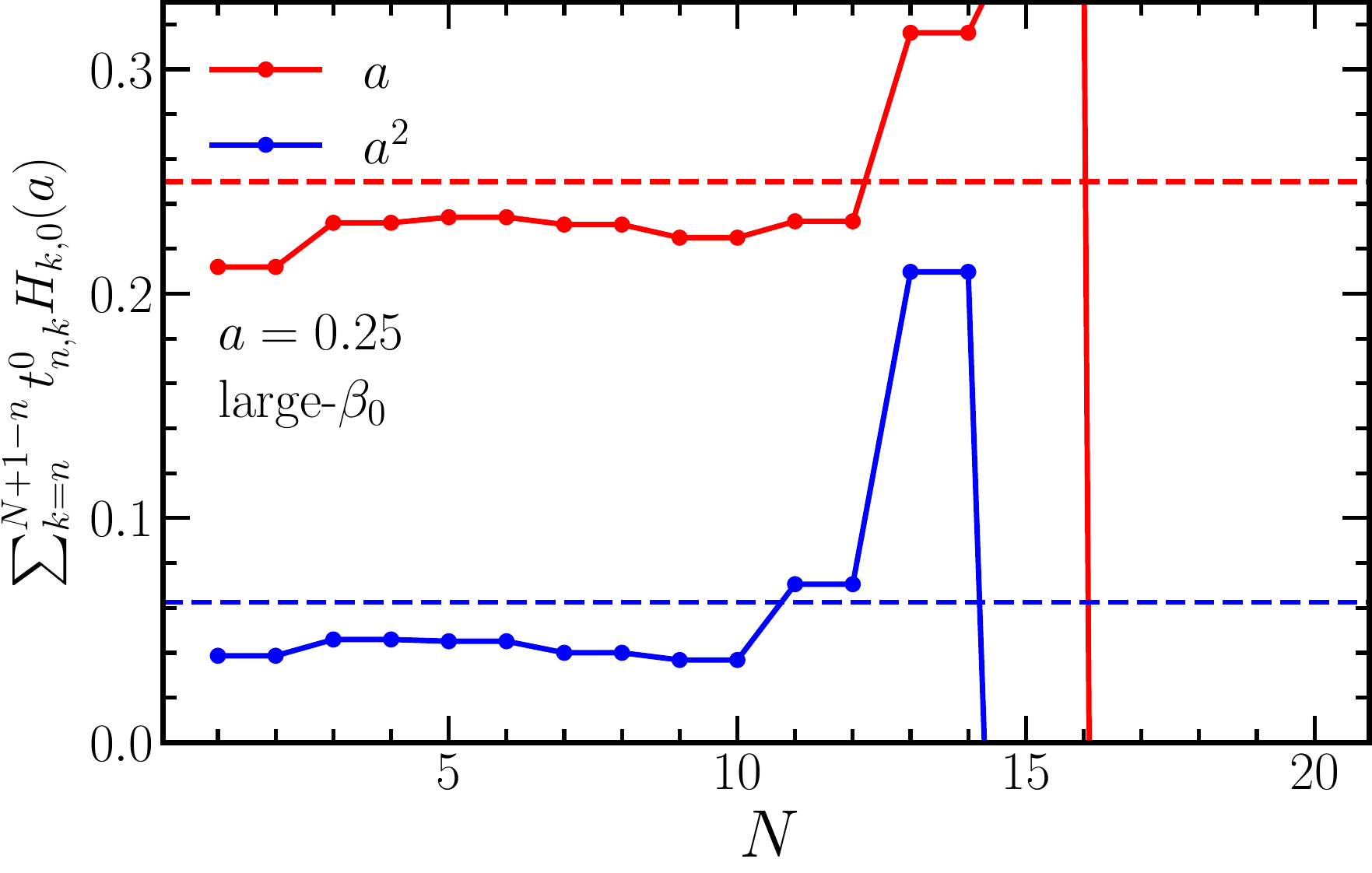}}
\subfigure[]{\includegraphics[width=.465\linewidth]{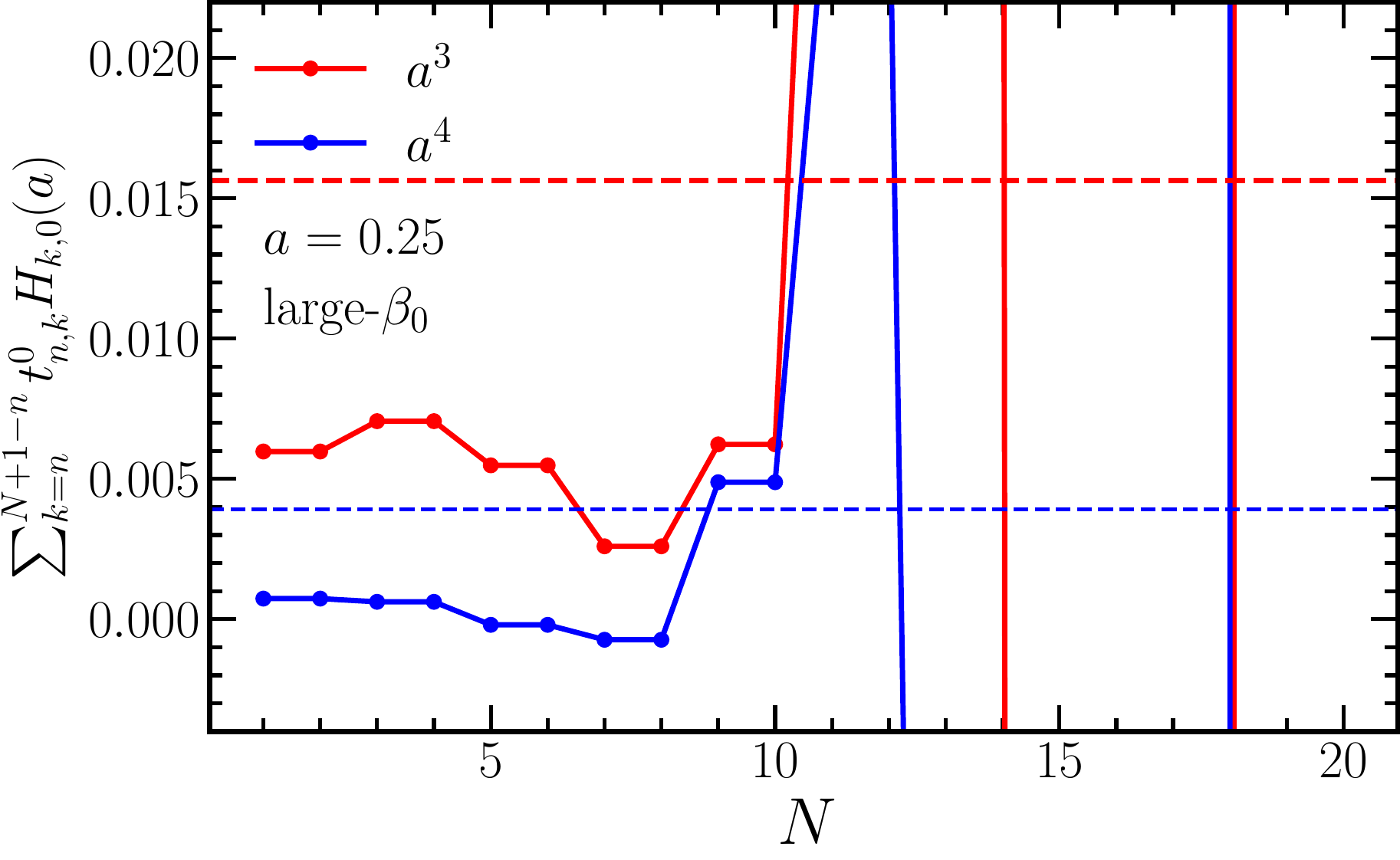}}
\caption{The series of Eq.~\eqref{eq:aninHk} for $\ell=0$: Expansion of $a^n$ in terms of the functions $H_{k,0}(a)$ for $a=0.25$.
We show results for $n=1,2$ in panel a and $n=3,4$ in panel b. Dashed horizontal lines represent the exact value for $a^n$.}
\label{fig:anFromHnl}
\end{figure*}

Let us now consider the transformation of a FOPT series $\sum^\infty_{n = 1} d_n a^n $ into a double series of $H_{n, \ell}(a)$ functions, see Eq.~\eqref{eq:FOPTtransform}.
It is a straightforward but rather cumbersome task to determine the coefficients of the expansion of a power $a^n$ in terms of the $H_{k, \ell}(a)$ functions given in
Eq.~\eqref{eq:aninHk}, but the expressions can also be determined in closed analytic form by inverting the triangular matrix of the $s^\ell_{n,k}$ coefficients.
The result reads
\begin{align}
\label{eq:tnkresults}
t^0_{n,k} & =\frac{(i\pi)^{k-n}(2-2^{k-n})\Gamma(k)}{\Gamma(k-n+1)\Gamma(n)}B_{k-n}\,,\\
t^{\ell \ge 1}_{n, k} & = \frac{(- 1)^{\ell}}{n-1}\times\left\{ \begin{array}{ll}
(k - 1) t_{n - 1, k - 1}^0 &\quad n + k \quad
\text{even}\\
- \ell\, t_{n - 1, k}^0 &\quad n + k \quad \text{odd} \nonumber
\end{array} \right.\,,
\end{align}
where $B_n$ are the Bernoulli numbers. It is now instructive to have a look at the asymptotic expression for these coefficients as $k\to\infty$.
Accounting for the factorial asymptotic growth of the Bernoulli numbers, see Eq.~\eqref{eq:Bernoulliasy}, the result reads
\begin{equation}
t^0_{n,k} \stackrel{k\to\infty}{\asymp} \frac{2i^{k-n}\Gamma(k)}{\Gamma(n)}(1-2^{n-k+1})\cos\!\biggl[\frac{(k-n)\pi}{2}\biggr]\,.
\end{equation}
This gives
\begin{equation}
\label{eq:anCIPTlimsup}
\limsup_{k\to\infty} \Bigl|t_{n,k}^{\ell} H_{k, \ell}(a)\Bigr|^{1/k} =
\infty\,.
\end{equation}
The rather interesting outcome is that the expansion of a simple power $a^n$ in terms of the $H_{n, \ell}(a)$ functions does not have any region of convergence and is factorially diverging for any value of $a$.
For illustration, we have displayed the expressions for $|t^\ell_{n,k}|^{1/k}$ as a function of $k$ for different values of $(n,\ell)$ in Fig.~\ref{fig:Largeb0T}.
So if we start from an absolutely convergent FOPT series, the resulting CIPT series will, {\it in general}, be divergent for any $a$.
The simplest example of an absolutely convergent FOPT series is that of a single term $a^n$. We have shown the $H_{k, \ell}$ expansions of $a^n$ truncated at order $N$ for $n=1,2,3,4$, $\ell=0$ and $a=0.25$
in Fig.~\ref{fig:anFromHnl}. We see that all the series look quite good and convergent at low orders, but they eventually diverge. Furthermore, in the range of orders $N$, where the series stabilize (and closely approach a particular value), the truncated series show a finite systematic discrepancy to the actual value of $a^n$ (indicated by the horizontal dashed lines). Hence, for a general absolutely convergent FOPT series the horizontal sums $\sum_{k = n+(\delta_{\ell,0}-1)}^\infty d_n t_{n,k}^{\ell}H_{k, \ell}(a)$ in Eq.~\eqref{eq:FOPTtransform} do not converge for any $n$ or $\ell$. The sum over $k$ of the vertical series values
$s_k=\sum_{n = 2-\delta_{\ell,0}}^{k+\delta_{\ell,0}} d_n t_{n,k}^{\ell}H_{k, \ell}(a)$
(which by themselves are finite due to the zeros of the array below the diagonal) is in general divergent as well, as we have demonstrated in the examples discussed in Sec.~\ref{sec:W1CIPT}. It is possible that the sum of the vertical series values $s_k$ is convergent as we have seen just above for the opposite transformation. However, this case is not the general one. For the application to $\tau$ hadronic spectral function moments, the latter case would arise in the situation that the Adler function series
of Eq.~\eqref{eq:AdlerD} were convergent. This situation does, however, not take place due to the unavoidable appearance of IR renormalons. It is also impossible that the divergent behavior of the coefficients $t_{n,k}^{\ell}$ as $k\to\infty$ may somehow compensate for a divergent behavior in a FOPT series yielding a convergent CIPT series for any value of $a$ as this would contradict
our findings from the beginning of this subsection.

In the discussion above we have considered the CIPT expansion in terms of $H_{n, \ell}$ functions for a single $\ell$. In phenomenological applications, linear combinations of $\ell$ always arise due to the property $W(1)=0$ already mentioned several times before. In this case, the same linear combinations of $s_{n,k}^{\ell}$ coefficients arise in the analogue of Eq.~\eqref{eq:Hninak}. The associated inverse $t_{n,k}$ coefficients, however, cannot be determined from the expressions in Eq.~\eqref{eq:tnkresults}, and no closed analytic expressions in analogy to Eqs.~\eqref{eq:tnkresults} can be given due to the many possible forms for the weight functions $W(x)$. Analytic expressions for definite values of $n$ and $k$ can, however, be obtained in a straightforward way.
We have checked for many physical weight functions, including the important kinematic weight function $W_{\tau}(x)=(1-x)^3(1+x)$ that the limit superior
for the resulting expressions $|t_{n,k}|^{1/k}$ as $k\to\infty$ is divergent just as for a single $\ell$. In particular, we do not find any cancellations of the kind discussed previously below Eqs.~\eqref{eq:limsupellp}, \eqref{eq:snkasy} and \eqref{eq:Hlimsup} for the case of physical weight functions that change the outcome of the discussion for a single $\ell$.

\subsection{On the Origin of the Behavior of the CIPT expansion}
The overall observation from the previous considerations is that using the CIPT expansion can turn a convergent FOPT series into a factorially divergent one, while the opposite can never ever happen. This means that CIPT is {\it a priori} inconsistent with the OPE and must be applied with great care, see Refs.~\cite{Benitez-Rathgeb:2022yqb,Benitez-Rathgeb:2022hfj}.
The diagnostic instrument to probe this behavior is to analyze the convergence properties of the series in Eq.~\eqref{eq:aninHk}, where a simple power $a^n$ is expressed as a series in $H_{k, \ell}(a)$ functions. This is the approach as to how the consistency of other expansion methods with the OPE, which are based on functions of the strong coupling, may be tested as well. If the series for a simple power $a^n$ is showing a factorial divergence or does not have a finite region of convergence, inconsistencies with the OPE may arise.
Even though this diagnostic tool is quite efficient and straightforward to apply, it would be instructive to have a more direct qualitative insight into which particular property of the $H_{n, \ell}(a)$ function is actually responsible for this behavior, as this knowledge may be highly useful in practice.
It is the purpose of this section to explore this question.

An essential difference between the asymptotic power sequence $\{a^n\}$ and the asymptotic sequences $\{H_{n, \ell}(a)\}$ is that the latter are nonuniform. The uniform convergence of series expansions in functions is an important aspect, as uniformity states that there is a certain level of global rapidity of the convergence valid for the whole interval for which the expansion is defined. The uniform convergence of the series expansion of a function $F(x)=\sum^\infty_{n = 1}f_n(x)$ in terms of the functions $f_n(x)$ is frequently a sufficient condition such that doing a linear operation on $F(x)$ is equivalent to doing this operation on the individual $f_n(x)$ and a subsequent summation. Uniform convergence is frequently assumed to be valid without proof in many phenomenological particle-physics applications as it is frequently also true. Even though the $H_{n, \ell}$ functions are bounded and series of the $H_{n, \ell}$ functions can be uniformly convergent, as we mentioned in the discussion below Eq.~(\ref{eq:CIPTlimsup}), the nonuniformity of them as asymptotic sequences $\{H_{n, \ell}(a)\}$ appears to disrupt
their convergence properties in a serious way when a convergent power series is transformed into a CIPT series.
\begin{figure}[t]
\centering
\includegraphics[width=0.44\linewidth]{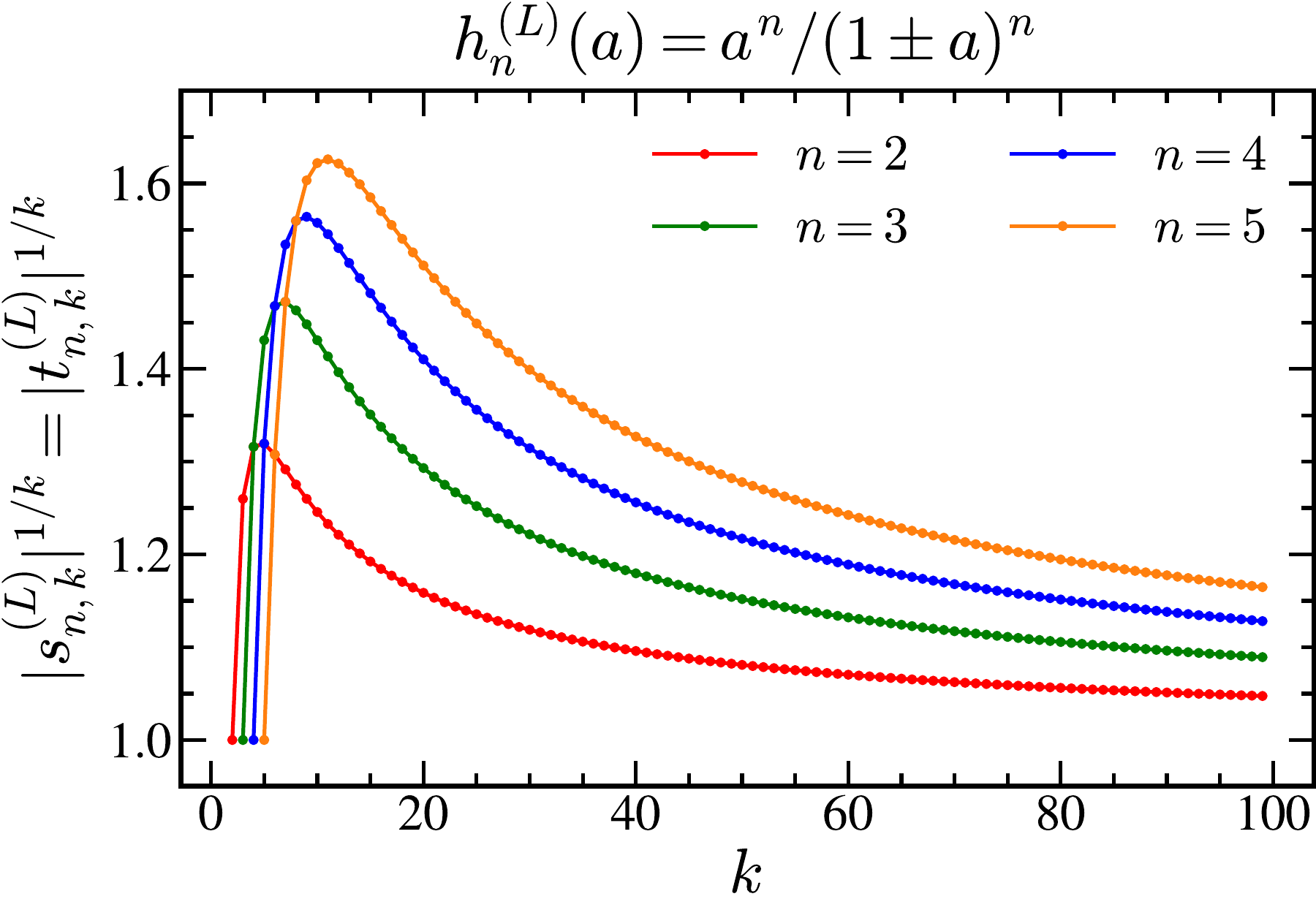} 
\caption{The $k$-th root of the $s$ and $t$ expansion coefficients defined on the RG-inspired model of Eq.~\eqref{eq:hnRGE}
for the values $L=\pm1$, to visualize Cauchy's root test. Red, green, blue, and orange correspond to the values $n=1,2,3,$ and $4$, respectively.
For this simple model, the absolute value of these coefficients is the same and hence are shown in a single plot.}
\label{fig:RunningST}
\end{figure}

As we have seen in Sec.~\ref{sec:W1CIPT}, this is because the ratio $H_{n+1,\ell} (a)/H_{n,\ell}(a)$ to be small in a region around $a=0$ requires the size of the region to shrink with $1/n$ when $n$ increases. This is related to the zeros in the functions $H_{n+1,\ell} (a)$ which approach $a=0$ like $1/n$ when
$n$ increases. We now explore whether these zeros, which are actually one reason why the CIPT expansion exhibits a quite rapid convergent appearance at lower orders, could be the origin of the problem. In the following we consider a number of simple toy asymptotic sequences to gain some more concrete insight concerning the answer of this question.

Let us discuss first the series transformation arising from a usual change of renormalization scale. Starting from a series in the powers $a^n=[a(s_0)]^n$ we can consider the reexpansion in terms of the powers $[a(\mu^2)]^n$ as an expansion in functions $h^{(L)}_n(a)$,
\begin{equation}
\label{eq:hnRGE}
h^{(L)}_n(a) = [a(\mu^2)]^n = \frac{a^n}{(1+a L)^n}\,,
\end{equation}
with $L=\log(\mu^2/s_0)$. The coefficients of the series $h^{(L)}_n(a)=\sum_{k=n}^\infty s^L_{n,k}a^k$ can be written down immediately and read
$s^L_{n,k}=\frac{(-L)^{k-n}\Gamma(k)}{\Gamma(n)\Gamma(k-n+1)}$. The coefficients of the inverse relation $a^n=\sum_{k=n}^\infty t^L_{n,k}h_k(a)$ are trivial to compute and read $t^L_{n,k}=(-1)^{k+n}s^L_{n,k}$. We have $\limsup_{k\to\infty} |s^L_{n,k}|^{1/k}=\limsup_{k\to\infty} |t^L_{n,k}|^{1/k}=|L|$ and both series are absolutely convergent for $|a|<1/|L|$. We have displayed $|s^L_{n,k}|^{1/k}=|t^L_{n,k}|^{1/k}$ for $|L|=1$ as a function of $k$ in Fig.~\ref{fig:RunningST}.
The $\{h^L_n(a)\}$ form a uniform asymptotic sequence since $|h^{(L)}_{n+1}(a)|/|h^{(L)}_{n}(a)|=|a/(1+a L)|$ does not depend on $n$. There are zeros in the functions $h^{(L)}_n(a)$ at $a=-1/L$, but they are independent of $n$ as well. So, as we can conclude from the Weierstrass double series Theorem~\ref{th:Weierstrassdouble}, using a fixed-order expansion at a different renormalization scale does not turn an absolute and uniformly convergent series into a nonconvergent one or may change the value of the series from the mathematical perspective, as long as $L$ is not too large. The usual approach of renormalization group improved calculations is to adapt $L$ such that the convergence happens in a rapid way.

\begin{figure*}[t]
\subfigure[]{\includegraphics[width=0.46\linewidth]{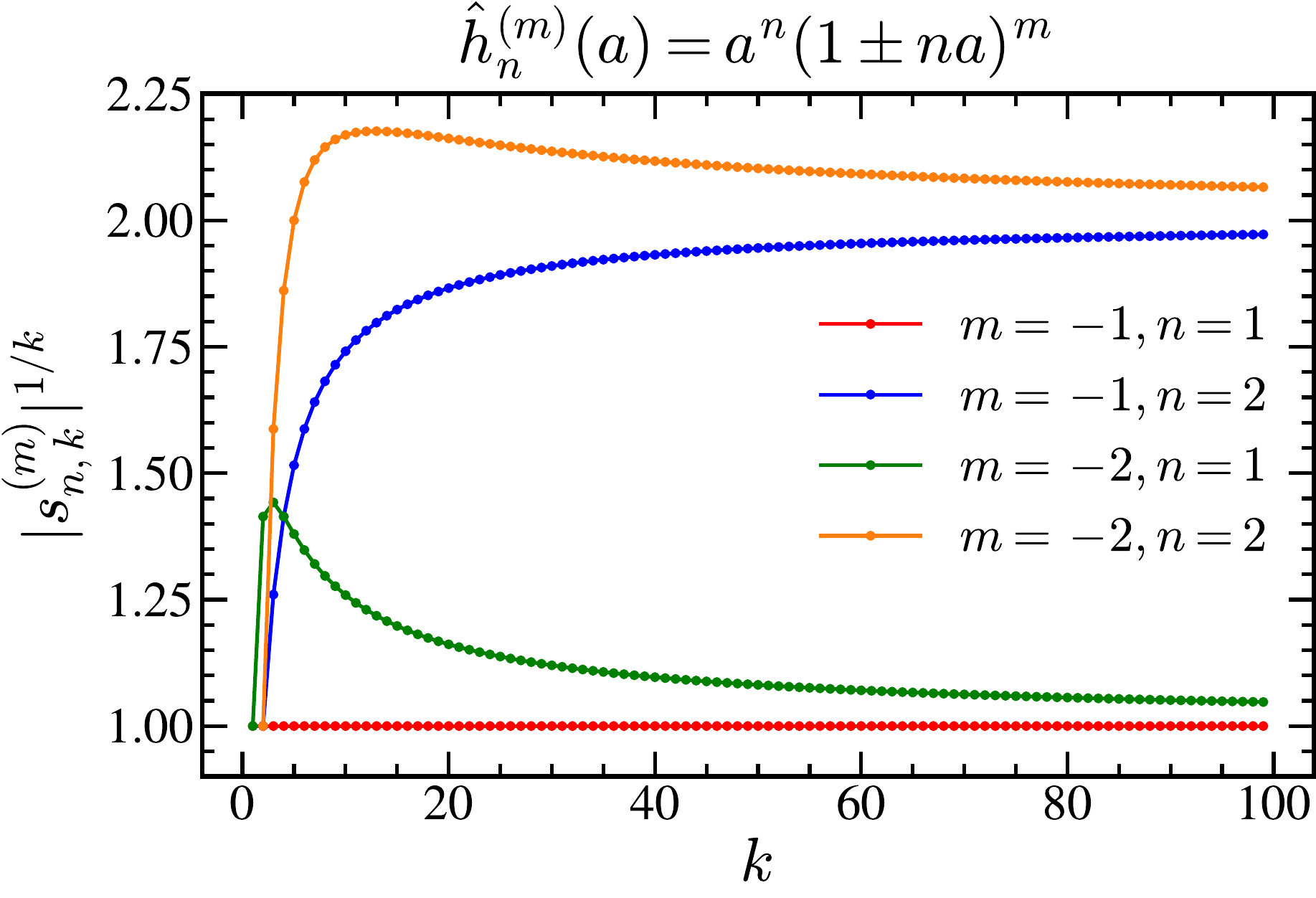}
\label{fig:ToyModelS}}
\subfigure[]{\includegraphics[width=0.45\linewidth]{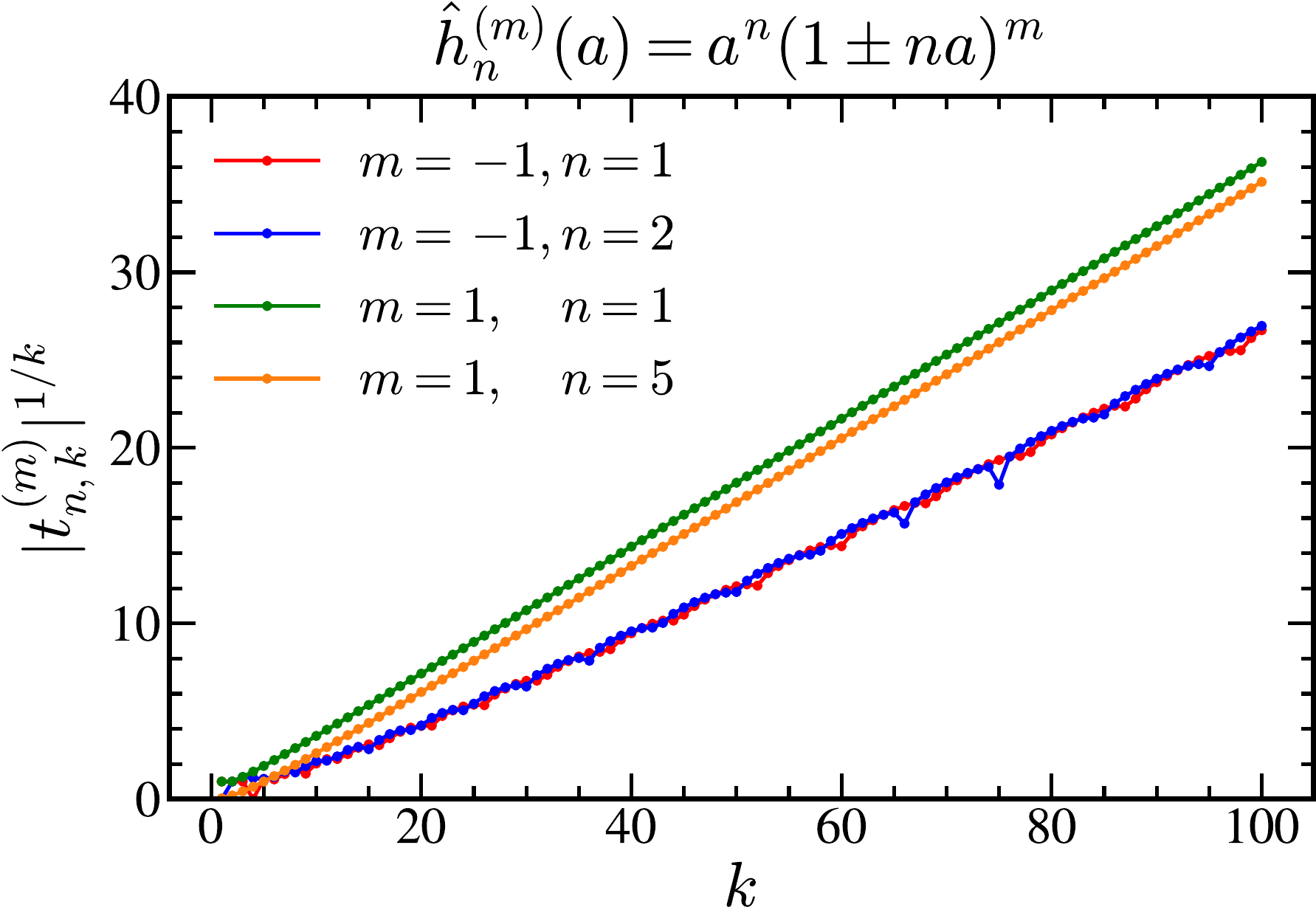}
\label{fig:ToyModelT}}
\caption{The $k$-th root of the $|s^{(m)}_{n,k}|$ (Panel a) and $|t^{(m)}_{n,k}|$ (Panel b) expansion coefficients defined on the model given in Eq.~\eqref{eq:hntoy1}
for the values $\xi=\pm1$. Panel a: red, and blue (red and green) use $m=-1$ ($n=1$) while green and orange (blue and orange) have $m=-2$ ($n=2$).
Panel b: red, and blue (red and green) use $m=-1$ ($n=1$) while green and orange have $m=-2$; green and orange correspond to $n=1$ and $5$, respectively.}
\label{fig:Toys}
\end{figure*}

Let us now consider the toy expansion functions
\begin{equation}
\label{eq:hntoy1}
\hat h_n^{(m)}(a) = a^n(1-\xi na)^m\,,
\end{equation}
with $m$ a positive or negative integer and $\xi$ some finite number.
We write the series of the $\hat h_n^{(m)}$ functions in powers of $a$ as $\hat h_n^{(m)}(a)=\sum_{k=n}^\infty s^{(m)}_{n,k}a^n$ and the series of $a^n$ in terms of the $\hat h_n^{(m)}$ functions as $a^n=\sum_{k=n}^\infty t^{(m)}_{n,k}\hat h_n^{(m)}(a)$
The $\{\hat h_n^{(m)}(a)\}$ form asymptotic sequences, but they are nonuniform due to the zeros or pole singularities at $a=1/(\xi n)$. It is obvious that a series in $\hat h_n^{(m)}$ functions for negative $m$ has bad properties since the singularities make it impossible to approximate any function that is continuous in some finite neighborhood of $a=0$. In Figs.~\ref{fig:ToyModelS} and ~\ref{fig:ToyModelT} we have displayed the values for $|s^{(-1)}_{n,k}|^{1/k}$ and $|t^{(-1)}_{n,k}|^{1/k}$ for $|\xi|=1$ as a function of $k$, clearly showing factorially diverging behavior of the \mbox{$t^{(-1)}_{n,k}$ coefficients} while the $s^{(-1)}_{n,k}$ are well-behaved. In this case, this does not come as a surprise.
However, for positive $m$ the situation appears not to be that bad due to the absence of the singularities and the emergence of zeros.
So, let us have a closer look at the case $m=1$. We have $s^{(1)}_{n,k}=\delta_{n,k}-\xi n\delta_{n+1,k}$. The series only has two terms and therefore trivially converges for all $a$. However, the coefficients of the inverse series read $t_{n,k}^{(1)}=\Gamma(k)\xi^{k-n}/\Gamma(n)$ with $t_{n,k}^{(1)}=0$ for $k<n$. We see immediately that this expansion is factorially divergent and does not have any finite region of convergence. For illustration we have displayed $|t^{(1)}_{n,k}|^{1/k}$ for $n=1,5$ as a function of $k$ in Fig.~\ref{fig:ToyModelT} as well.
The interesting insight gained by the toy expansion functions $\hat h_n^{(m)}(a)$ is that zeros (for positive $m$) as well as singularities (for negative $m$) yield the same badly diverging behavior for the $t_{n,k}^{(m)}$ coefficients.

We have tested a number of other toy expansion functions forming nonuniform asymptotic sequences due to zeros or singularities approaching $a=0$ for large $n$, all leading to factorially diverging $t_{n,k}$ coefficients. Even though we do not claim that out findings provide
the exact mathematical specification under which the $t_{n,k}$ coefficients factorially diverge as $k\to\infty$, we believe that the presence (or absence) of the uniformity property of the asymptotic sequences $\{h_n(a)\}$ plays an essential role for the expansion functions to be consistent with the OPE. While it is obvious that expansion functions, where the nonuniformity is caused by pole-type singularities or other types of nonanalytic properties, are not suitable for phenomenological applications, the certainly surprising aspect is that even zeros, which may appear to have advantages for a rapid convergence at first sight, can have the same bad effect.

\section{Generalization to full QCD}
\label{sec:WxellQCD}

We conclude this article by considering the case of full QCD, where all terms of the QCD $\beta$-function are accounted for.
Even though it appears quite unreasonable to argue that our observations concerning the asymptotic sequences $\{a^n\}$ and $\{ H_{n,\ell}(a)\}$ in the large-$\beta_0$ approximation may not apply in full QCD, it is worth having a closer look. However, in contrast to the large-$\beta_0$ approximation, we cannot obtain fully analytic results. We therefore rely on numerical studies and evidences rather than strict proofs. Still some closed formulas can be presented in the C-scheme for the strong coupling $\alpha_s$~\cite{Boito:2016pwf}. The C-scheme uses that only the one- and two-loop coefficients of the QCD $\beta$-function are renormalization-scheme invariant. It is therefore possible to adopt a scheme for $\alpha_s$ where the QCD $\beta$-function takes the all-order form
\begin{equation}
\label{eq:betaC0}
\frac{\dd a(\mu^2)}{\dd\ln \mu^2} \,=\,
-\frac{[a(\mu^2)]^2}{1-2\,\hat b_1 a(\mu^2)} \,,
\end{equation}
where $\hat b_1=\beta_1/(2\beta_0^2)$ and $\beta_1=102-38\,n_f/3$. We furthermore demand that the C-scheme strong coupling differs from the common $\overline{\rm MS}$ strong coupling by terms of order $\alpha_s^3$ and higher, which unambiguously fixes the C-scheme strong coupling to all orders. Furthermore, the QCD scale $\Lambda_{\rm QCD}$ agrees with the one in the $\overline{\rm MS}$ scheme.\footnote{In Ref.~\cite{Boito:2016pwf} this strong coupling scheme was called the $(C=0)$-scheme and it is numerically very close to the $\overline{\rm MS}$ strong coupling, see also the Appendix of Ref.~\cite{Benitez-Rathgeb:2022yqb}.} For $\hat b_1=0$ the C-scheme agrees with the large-$\beta_0$ approximation.
In the following we gather sufficient evidence that all qualitative insights obtained in the previous sections in the large-$\beta_0$ approximation are true as well in full QCD.

\begin{figure*}
\subfigure[]{\includegraphics[width=0.45\linewidth]{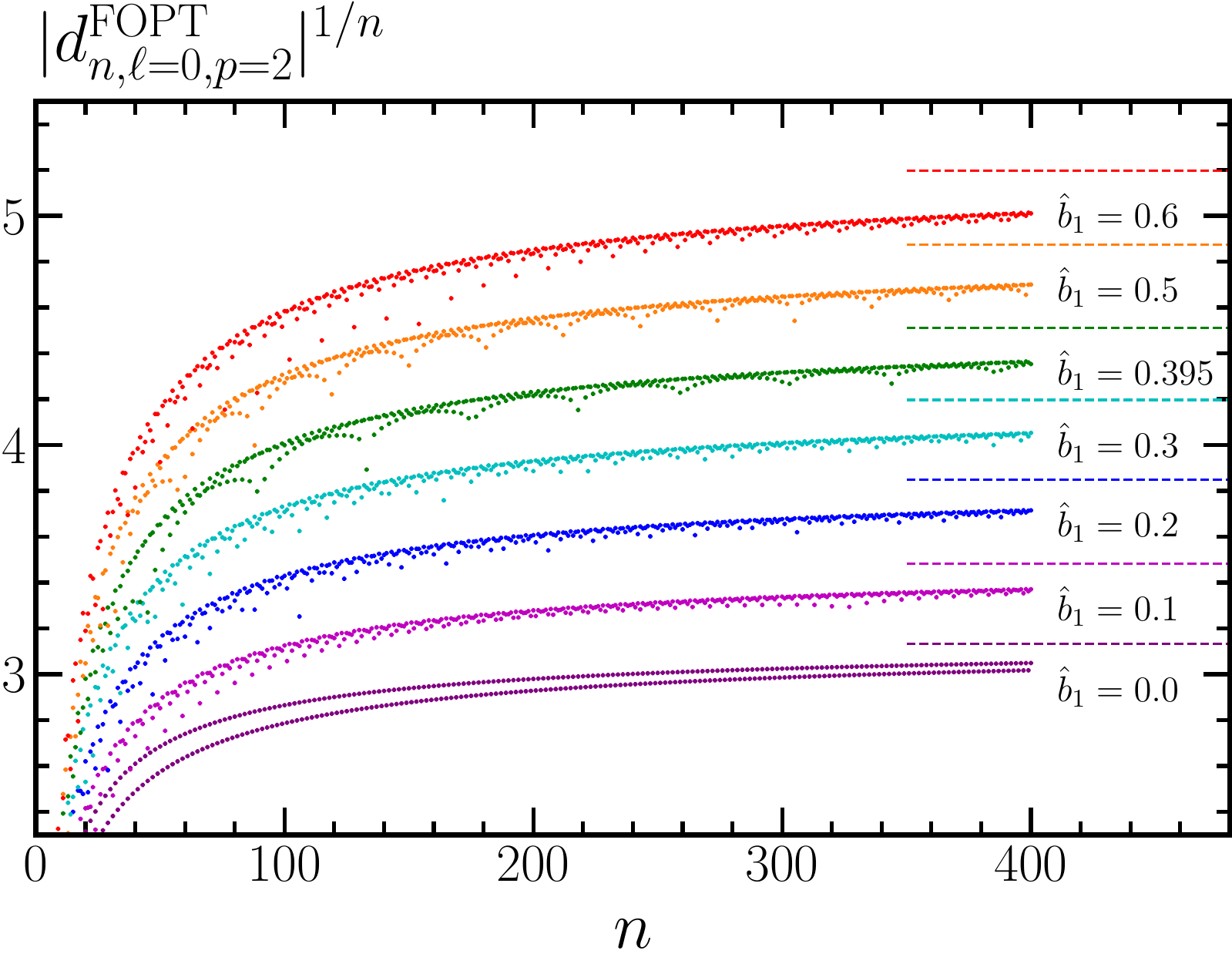}
\label{fig:FOPTRoot1}}
\subfigure[]{\includegraphics[width=0.455\linewidth]{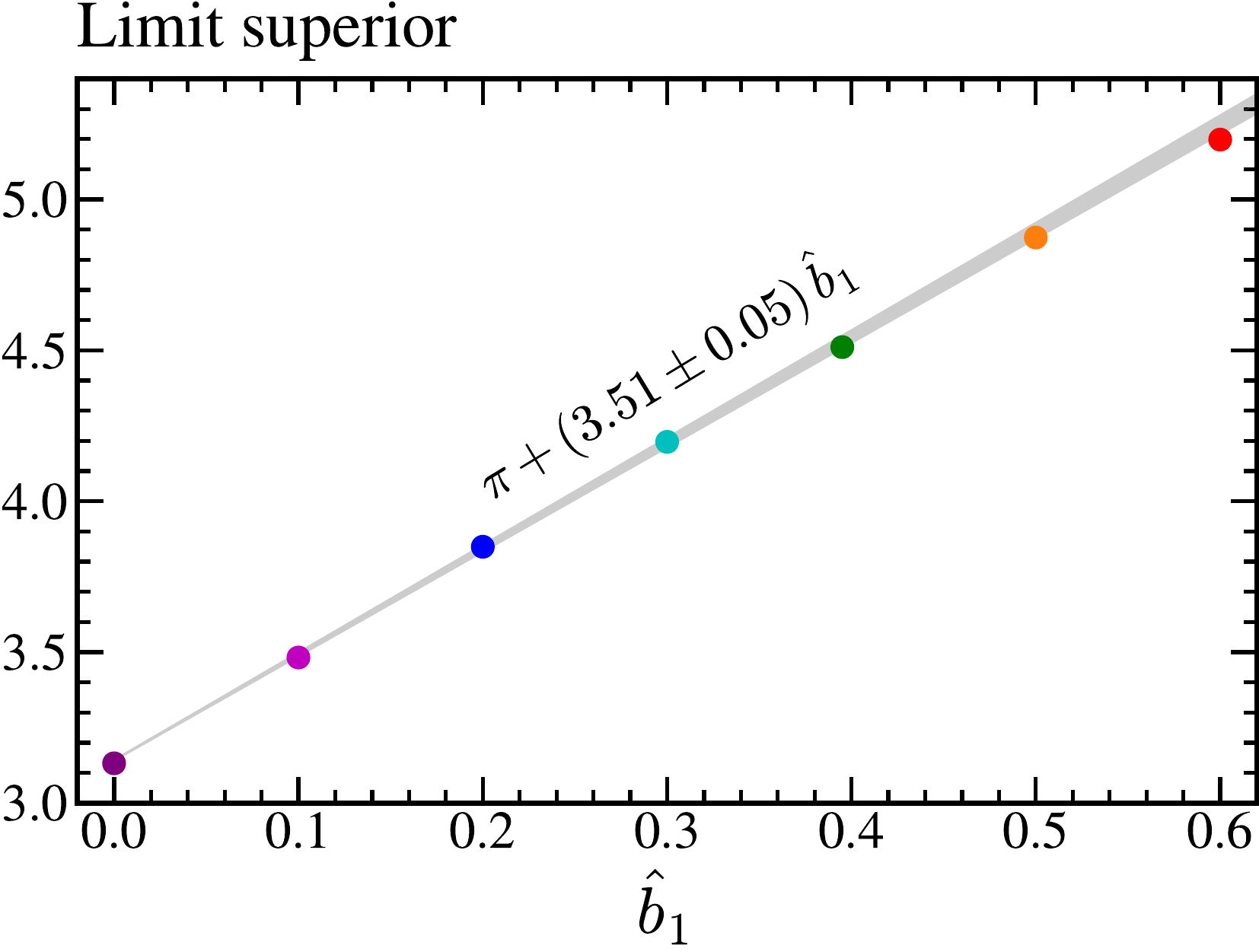}
\label{fig:FOPTRoot2}}
\caption{Left panel illustrates the root test for $\delta^{(0)}_{\text{FOPT},\ell,p}$ with $\ell=0$ and $p=2$ for various values of $\hat{b}_1$. The horizontal dashed lines represent our estimates for the limit superior for this analysis, which are shown to have a linear dependence on $\hat{b}_1$ in the right panel. The gray area represents
the result of Eq.~(\ref{eq:fitforconvradius}).}
\label{fig:FOPTRoot}
\end{figure*}

\subsection{Spectral Function Moment Series in FOPT and CIPT}
\label{sec:W1QCD}

We start by considering the full QCD generalization of the spectral function moment analysis of Sec.~\ref{sec:W1largeb0}. The full QCD generalization (in the C-scheme) of the single pole Borel function in the large-$\beta_0$ approximation corresponding to an OPE matrix-element corrections of the form $\mbox{const.}\times\langle{\cal O}_p\rangle/s_0^p$ in the Adler function
reads $B_{p,\hat b_1}(u)=1/(p-u)^{1+2p\hat{b}_1}$ (with $p=2,3,\ldots$) and consistency with the OPE again demands that the resulting perturbative series of the spectral function moment $\delta_{\ell\neq p}^{(0)}$ have a finite region of convergence. The resulting coefficients in the Adler function series of Eq.~\eqref{eq:AdlerD} read
\begin{align}
\label{eq:cn1QCD}
\bar{c}_{n,1,p} &\,= \frac{\Gamma(n+2p\hat{b}_1)}{p^{n+2p\hat{b}_1}\Gamma(1+2p\hat{b}_1)}\,,\\
\nonumber
\bar c_{n ,k\ge 2,p} &=\, -\frac{1}{k} \sum_{j = k-1}^{n - 1} (2 \hat b_1)^{n - j - 1}
j \, \bar c_{j, k - 1,p} \,.
\end{align}

\begin{figure*}
\subfigure[]{\includegraphics[width=0.45\linewidth]{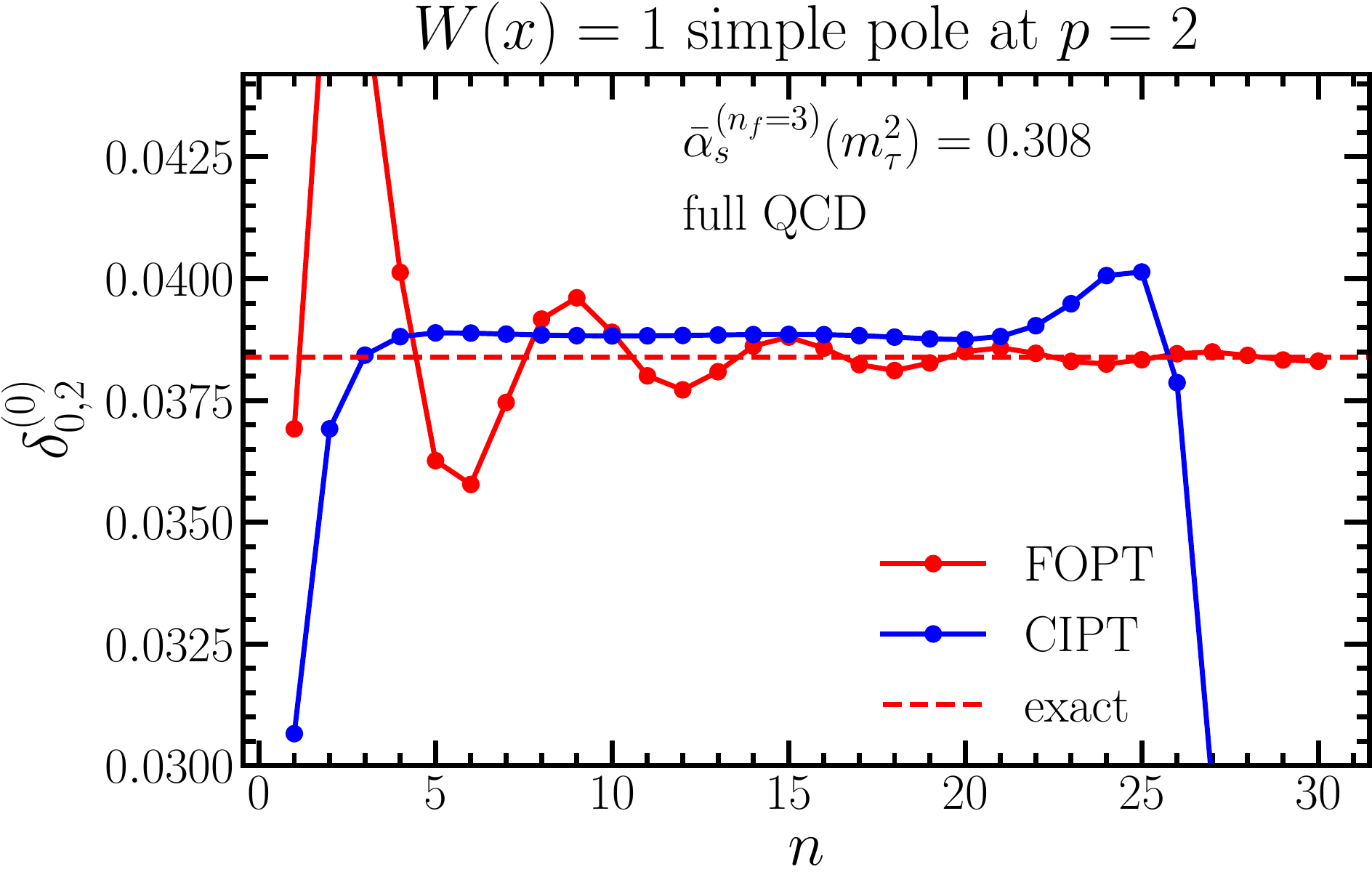}}
\subfigure[]{\includegraphics[width=0.45\linewidth]{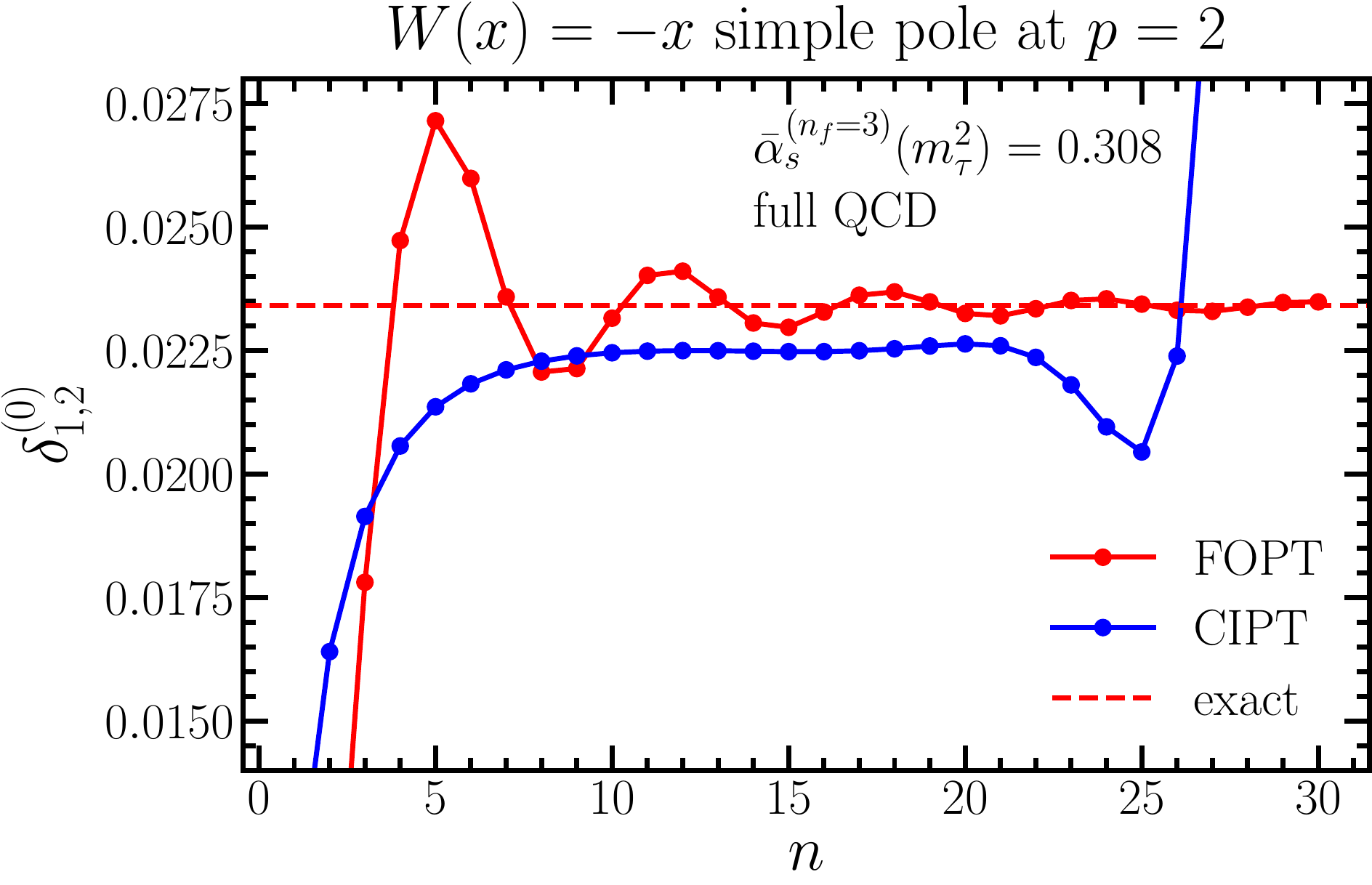}}
\caption{Same as in Fig.~\ref{fig:FOPTvsCIPT} but for the C-scheme with $\hat{b}_1=32/81\simeq 0.395$ and where the value of the strong coupling in the
$\overline{\rm MS}$ is $\alpha_s(m_\tau^2)=0.315$.}
\label{fig:FOPTvsCIPTCScheme}
\end{figure*}

Let us first analyze the FOPT series coefficients $d^{\text{FOPT}}_{n, \ell, p \not{=} \ell}$. We have derived them for many different choices for $\ell\neq p$ and find that the
sequences $|d^{\text{FOPT}}_{n, \ell, p \not{=} \ell}|^{1 / n}$ in $n$ are indeed consistent with sequences that converge for all cases. We find that the limit superior approached by these series does not depend on the values of $\ell$ and $p$, but it does on the value of $\hat b_1$. We also find that this limit superior agrees with the one obtained for the series of $a_{\pm}\equiv a(-s_0\pm i0)$ in powers of $a=a(s_0)$.
In Fig.~\ref{fig:FOPTRoot1} we have displayed $|d^{\text{FOPT}}_{n,\ell=0,p=2}|^{1/n}$ for orders up to $n=400$ for different values of $\hat b_1$. Carrying out a quadratic fit in $1/n$ we determined an accurate estimate for the limit superior which depends linearly on $\hat b_1$ to very good approximation. We have displayed the outcome of this analysis for $|d^{\text{FOPT}}_{n,\ell=0,p=2}|^{1/n}$ in Fig.~\ref{fig:FOPTRoot2} together with a fit function quantifying the coefficient of $\hat b_1$. The outcome for a combined analysis using the coefficients $d^{\text{FOPT}}_{n, \ell, p}$ for $\ell=0,1,2,3,4$ and $p=1,2,3,4$ with $\ell\neq p$
and including also the series for $a_\pm$ for orders $250\le n\le 400$ yields the result
\begin{align}
\label{eq:fitforconvradius}
\underset{n \rightarrow \infty}{\limsup}\,\bigl| d^{\rm FOPT}_{n, \ell, p \not{=} \ell} \bigr|^{1 / n} \, = \,
\pi + (3.51 \pm 0.05)\,\hat b_1\,,
\end{align}
where the quoted uncertainty represents the values covered by the individual analyses.
For $\hat b_1=0$ we obtain $\pi$ with very high accuracy, consistent with our analytic results in the \mbox{large-$\beta_0$} approximation in Eq.~\eqref{eq:limsupellp}.
Our results reconfirm the convergence proof in Ref.~\cite{Benitez-Rathgeb:2022yqb} based on the renormalon calculus already mentioned at the end of Sec.~\ref{sec:W1FOPT} and can be used to determine the radius of convergence of the FOPT series $\delta_{{\FOPT}, \ell}^{(0)}$ [and the expansions of $a_\pm$ in powers of $a=a(s_0)$]. We have displayed the convergent FOPT moment series for the cases $(\ell,p)=(0,2)$ and $(1,2)$ in Fig.~\ref{fig:FOPTvsCIPTCScheme} as the red dots. We note that we have also checked the FOPT series coefficients for physical weight functions with $W(1)=0$. In analogy to our analytic findings for the large-$\beta_0$ approximation we found that there is a significant cancellation among the coefficients $d^{\text{FOPT}}_{n, \ell, p \not{=} \ell}$, which does, however, not affect the limit superior. Also for physical weight functions the result given in Eq.~\eqref{eq:fitforconvradius} applies.

For $n_f=3$, which is relevant for hadronic $\tau$ decays, we have $\hat b_1=32/81\simeq 0.395$ which yields $\limsup_{n\to\infty}|d^{\rm FOPT}_{n, \ell=0, p=2}|^{1 / n}=4.528\pm 0.020$
giving \mbox{$\alpha_s(s_0)=0.3083 \pm 0.0013$}
as the convergence radius in the \mbox{C-scheme}. This corresponds to $\alpha_s(s_0)=0.3151\pm 0.0012$
in the $\overline{\rm MS}$ scheme which for $s_0=m_\tau^2$ is right within the world average of the $\overline{\rm MS}$ strong coupling at the $\tau$ lepton scale \mbox{$\alpha_s(m_\tau^2)=0.312\pm 0.015$}. This interesting fact was already pointed out many years ago (albeit with much lower precision) by Pich and Le Diberder in Ref.~\cite{LeDiberder:1992jjr}. So using FOPT may have issues as well.
In practice, however, the behavior of the FOPT series at low orders is still quite good for all values close to the convergence radius due to the cancellation among the $d^{\text{FOPT}}_{n, \ell, p}$ coefficients for physical weight functions mentioned above.
It is furthermore easy to check that the FOPT series behavior at intermediate orders is similar regardless whether $\alpha_s(m_\tau^2)$ is chosen slightly above or below the convergence radius.
At this point we remind the reader that here we discuss the behavior of FOPT for a particular contribution in the hadronic $\tau$ spectral function moment series. In full phenomenological applications, the FOPT series is asymptotic due to the unavoidable appearance of coefficients $d^{\text{FOPT}}_{n, \ell, p}$ for $\ell=p$.
Our finding, however, implies that high-precision determinations of the strong coupling from hadronic $\tau$ spectral function moments need to be interpreted with some care.

\begin{figure*}[t!]
\centering
\subfigure[]{\includegraphics[width=.405\linewidth]{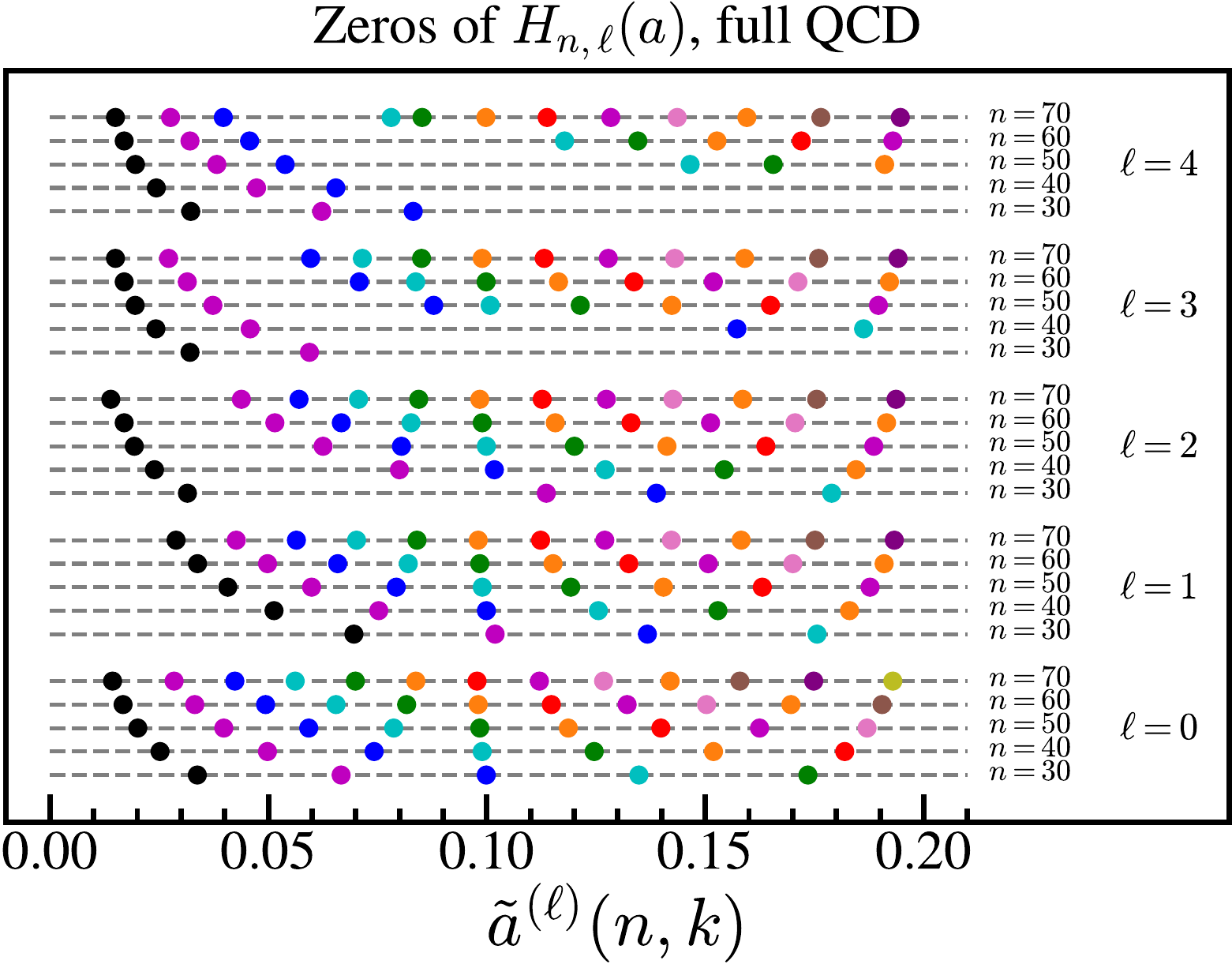}
\label{fig:ZerosHnl1FullQCD}}
\subfigure[]{\includegraphics[width=.463\linewidth]{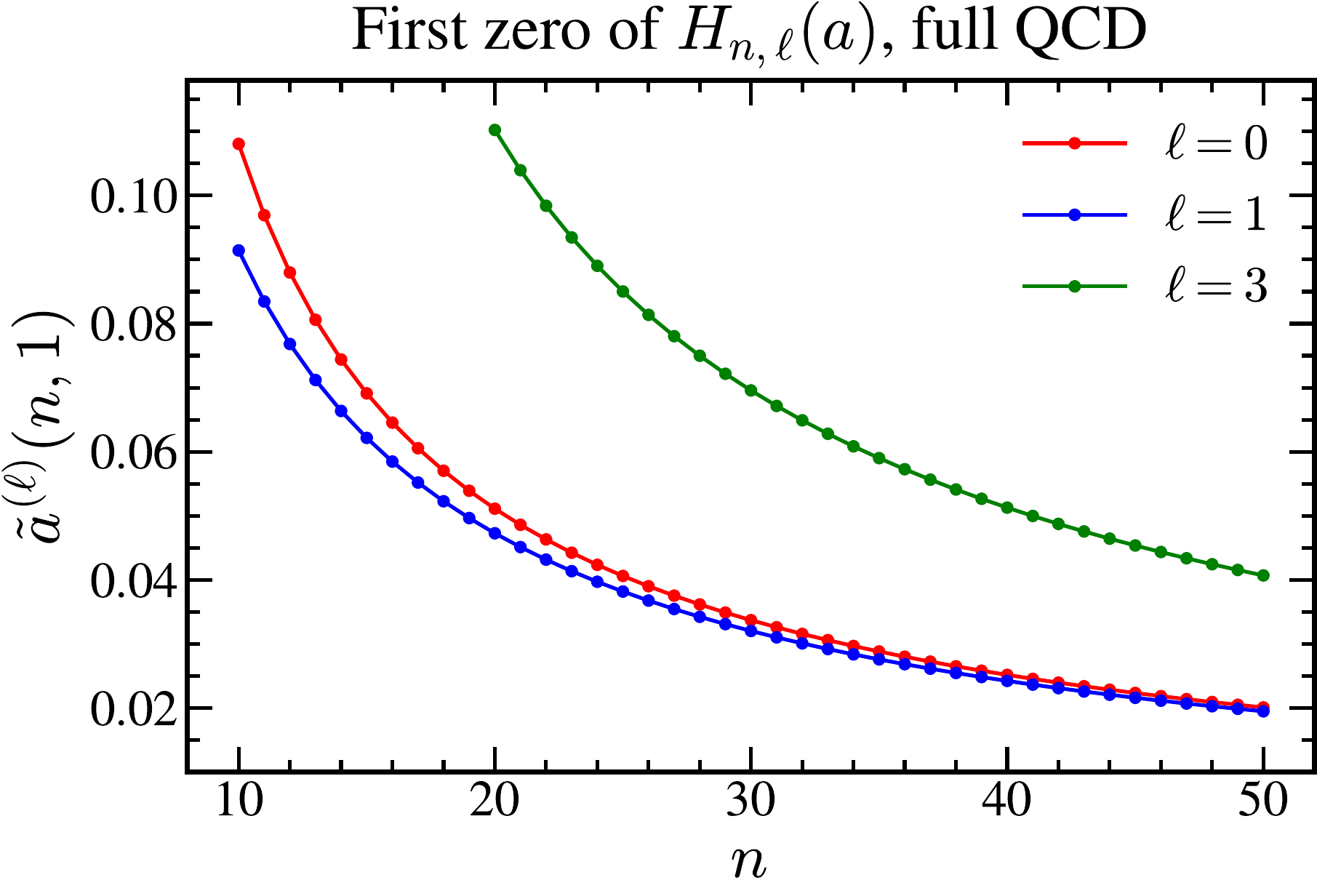}
\label{fig:ZerosHnl2FullQCD}}
\caption{Same as Fig.~\ref{fig:ZerosHnl} for the C-scheme with $\hat{b}_1=0.395$ in full QCD.}
\label{fig:ZerosHnlQCD}
\end{figure*}

The CIPT expansion functions $H_{n, \ell}(a)$ in full QCD have been determined analytically in Ref.~\cite{Hoang:2020mkw}\footnote{We take the opportunity to point out a typo in Ref.~\cite{Hoang:2020mkw}; Eq.~(3.13) should read $\tilde{H}(n,0,-1)=-i/\pi(-t_0)^n\ln(t_+/t_-)$. So the result on the RHS given in Ref.~\cite{Hoang:2020mkw} is correct, but it belongs to a different function than shown on the LHS.} as an infinite sum for positive real $a$. In the C-scheme the sum terminates due to the form of the $\beta$-function and the integral representation for the $H_{n, \ell}(a)$ functions can be written as
\begin{equation}
H_{n, \ell} (a) = \frac{i}{\pi} e^{- \frac{\ell}{a}} a^{- 2 \ell \hat{b}_1}\!\!
\int_{t_-}^{t_+}\!\! \dd t \biggl( 1 + \frac{\hat{b}_1}{t} \biggr) e^{- 2 \ell
t} (- 2 t)^{- 2 \ell \hat{b}_1 - n} \,.
\end{equation}
The integral reduces to Eq.~\eqref{Hndefb0} in the large-$\beta_0$ approximation, when $\hat b_1=0$.
The results can be given in closed form in terms of $a_\pm=a(-s_0\pm i0)$
and read
\begin{align}
\label{eq:H10fullQCDanalytic}
\!\!\!\!\!\!\!\! H_{1,0}(a)=& \frac{1}{2\pi i}\biggl[2\hat{b}_1(a_+-a_-)+\log\biggl(\frac{a_-}{a_+}
\biggr)\biggr],\\
\label{eq:Hn0fullQCDanalytic}
\!\!\!\!\!\!\!\! H_{n \geq 2, 0}(a)=& \frac{1}{2\pi i}\Biggl[\frac{2\hat{b}_1(a^n_+- a^n_-)}{n}-\frac{a^{n-1}_+-a^{n-1}_-}{n-1}\Biggr],\\
\label{eq:HnellfullQCDanalytic}
\!\! H_{n,\ell>0}(a)=& \frac{\ell^{n+2\ell\hat{b}_1-1}}{a^{2\ell\hat{b}_1}}e^{-\frac{\ell}{a}}\!\Biggl[\frac{n}{\Gamma(n+2\ell\hat{b}_1-1)}
+e^{-i\pi\hat{b}_1\ell}h_{n,\ell}(a_+,\hat{b}_1)-e^{i\pi\hat{b}_1\ell}h_{n,\ell}(a_-,\hat{b}_1)\!\Biggr],\!\!\!\!
\end{align}
where
\begin{equation}
\label{eq:littleh}
h_{n, \ell} (a, \hat{b}_1) \equiv \frac{(- 1)^{n + 1}}{2 \pi i} \Biggl[ n\,
\Gamma \biggl( - n - 2 \hat{b}_1 \ell, - \frac{\ell}{a} \biggr) -
e^{\frac{\ell}{a}} \biggl( - \frac{a}{\ell} \biggr)^{\!\!n + 2 \hat{b}_1 \ell}\,\Biggr].
\end{equation}
Since the first two terms in the expansion of the strong coupling at some scale in terms of powers of the strong coupling at another scale in full QCD and the large-$\beta_0$ approximation agree, the ratio of the $H_{n, \ell}$ functions defined in Eq.~\eqref{Hndef} in full QCD to the ones in the large-$\beta_0$ approximation approach unity in the limit $a\to 0$. It is therefore obvious that the $\{H_{n, \ell}(a)\}$ also represent asymptotic sequences in full QCD. We also find that the $H_{n, \ell}$ functions again exhibit zeros along the positive real $a$ axis approaching zero as $n\to\infty$.
In Fig.~\ref{fig:ZerosHnlQCD} we display the positive zeros of the $H_{n, \ell}$ functions in full QCD for a large number of $\ell$ and $n$ values for illustration in analogy to the large-$\beta_0$ approximation shown in Fig.~\ref{fig:ZerosHnl}. This shows that the asymptotic sequences $\{H_{n, \ell}(a)\}$ in full QCD are nonuniform as well.
Using relation~\eqref{eq:gammalargez} it is straightforward to derive the asymptotic expressions for the $H_{n,\ell>0}(a)$ as $n\to\infty$ for positive real $a$ yielding
\begin{equation}
H_{n,\ell}(a)\stackrel{n\to\infty}{\asymp}e^{\ell\bigl(\frac{\mathrm{Re}(a_+)}{|a_+|}-\frac{1}{a}\bigr)}
\frac{(|a_+|/a)^{2\ell\hat{b}_1}|a_+|^n}{n\pi}\biggl[2\hat{b}_1\sin(\theta_n)-\frac{\sin(\theta_{n-1})}{|a_+|}\biggr],
\end{equation}
with $\theta_n\equiv(2\ell\hat{b}_1+n)\arctan[\mathrm{Im}(a_+)/\mathrm{Re}(a_+)]-\ell\,\mathrm{Im}(a_+)/|a_+|^2$.
This allows us to also determine the limit superior of $|H_{n,\ell}(a)|^{1/n}$ as $n\to\infty$ in full QCD, which is surprisingly simple,
\begin{equation}
\label{eq:HlimsupQCD}
\limsup_{n\to\infty} |H_{n,\ell}(a)|^{1/n}=|a_+|
\,.
\end{equation}

As for the large-$\beta_0$ approximation, we can now see analytically that the CIPT series $\delta_{{\CIPT}, \ell}^{(0)}$
of Eq.~\eqref{eq:deltaCIPTdef} with the coefficients in Eq.~\eqref{eq:cn1QCD} diverge for any value of $a$,
\begin{equation}
\label{eq:CIPTlimsupQCD}
\limsup_{n\to\infty} |\bar{c}_{n,1} H_{n,\ell}(a)|^{1/n} = \infty\,.
\end{equation}
We have displayed the divergent CIPT moment series for the cases $(\ell,p)=(0,2)$ and $(1,2)$ in Fig.~\ref{fig:FOPTvsCIPTCScheme} as the blue dots, showing, just like in the large-$\beta_0$ approximation, the apparent convergence of the CIPT series at intermediate orders and the discrepancy to the value of the convergent FOPT series.

\begin{figure*}
\subfigure[]{\includegraphics[width=0.455\linewidth]{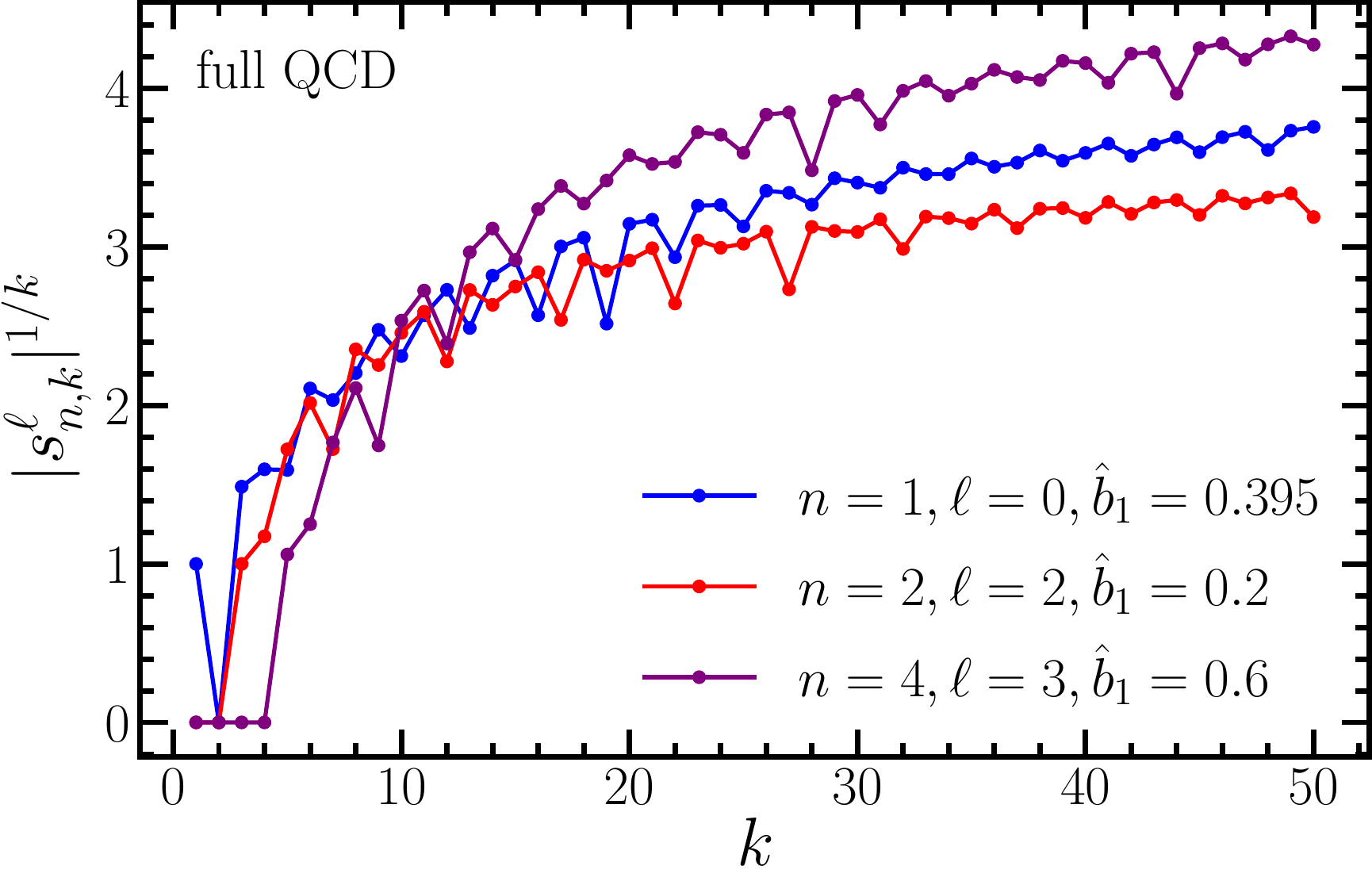}
\label{fig:SFullQCD}}
\subfigure[]{\includegraphics[width=0.47\linewidth]{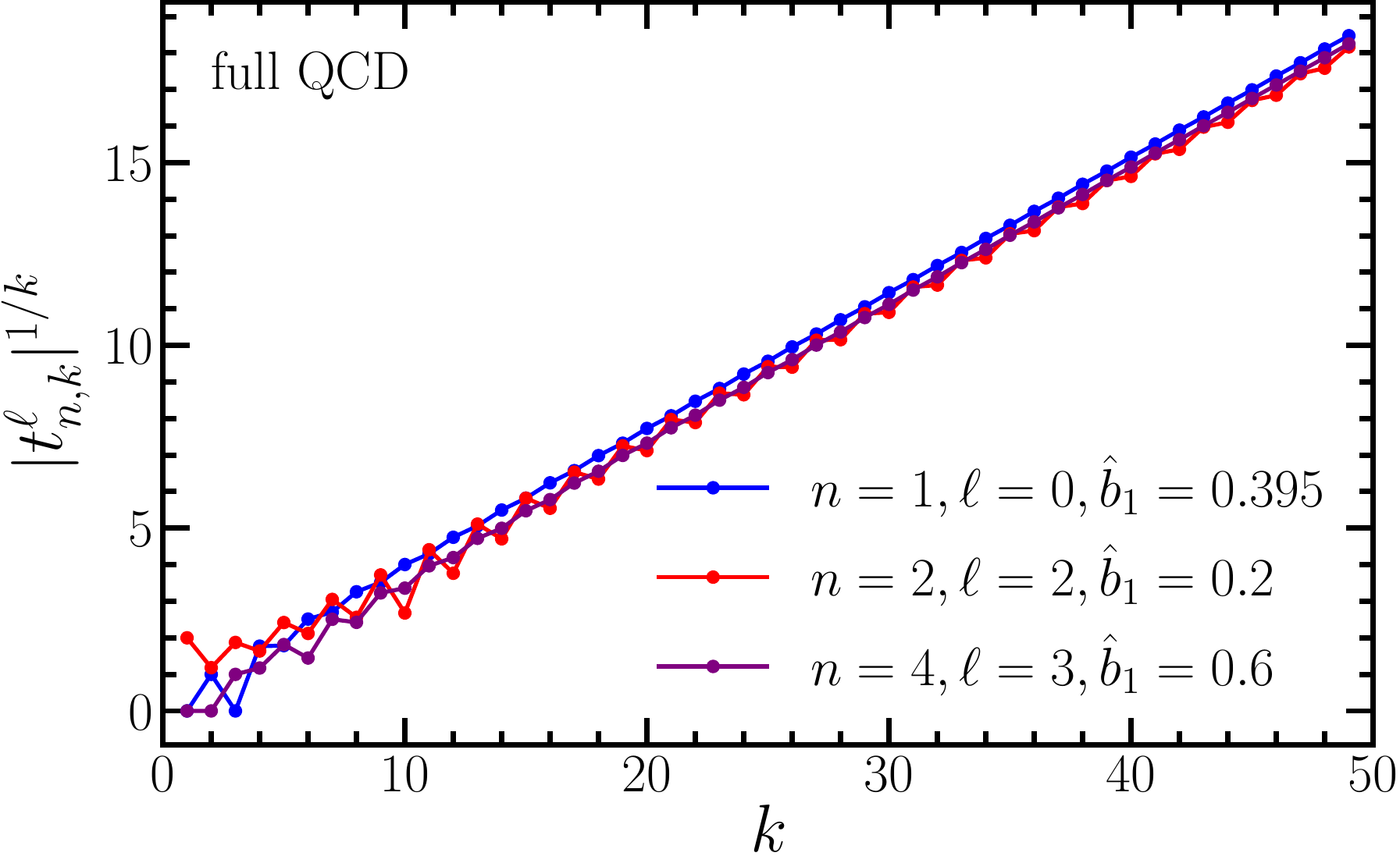}
\label{fig:TFullQCD}}
\caption{Same as Fig.~\ref{fig:ST} in full QCD, using the C-scheme for a number of scenarios: blue, red and purple correspond to $(n,\ell, \hat b_1)=(1,0, 0.395)$,
$(2,2, 0.2)$, and $(4,3, 0.6)$, respectively.}
\label{fig:STFullQCD}
\end{figure*}

\begin{figure*}
\subfigure[]{\includegraphics[width=0.485\linewidth]{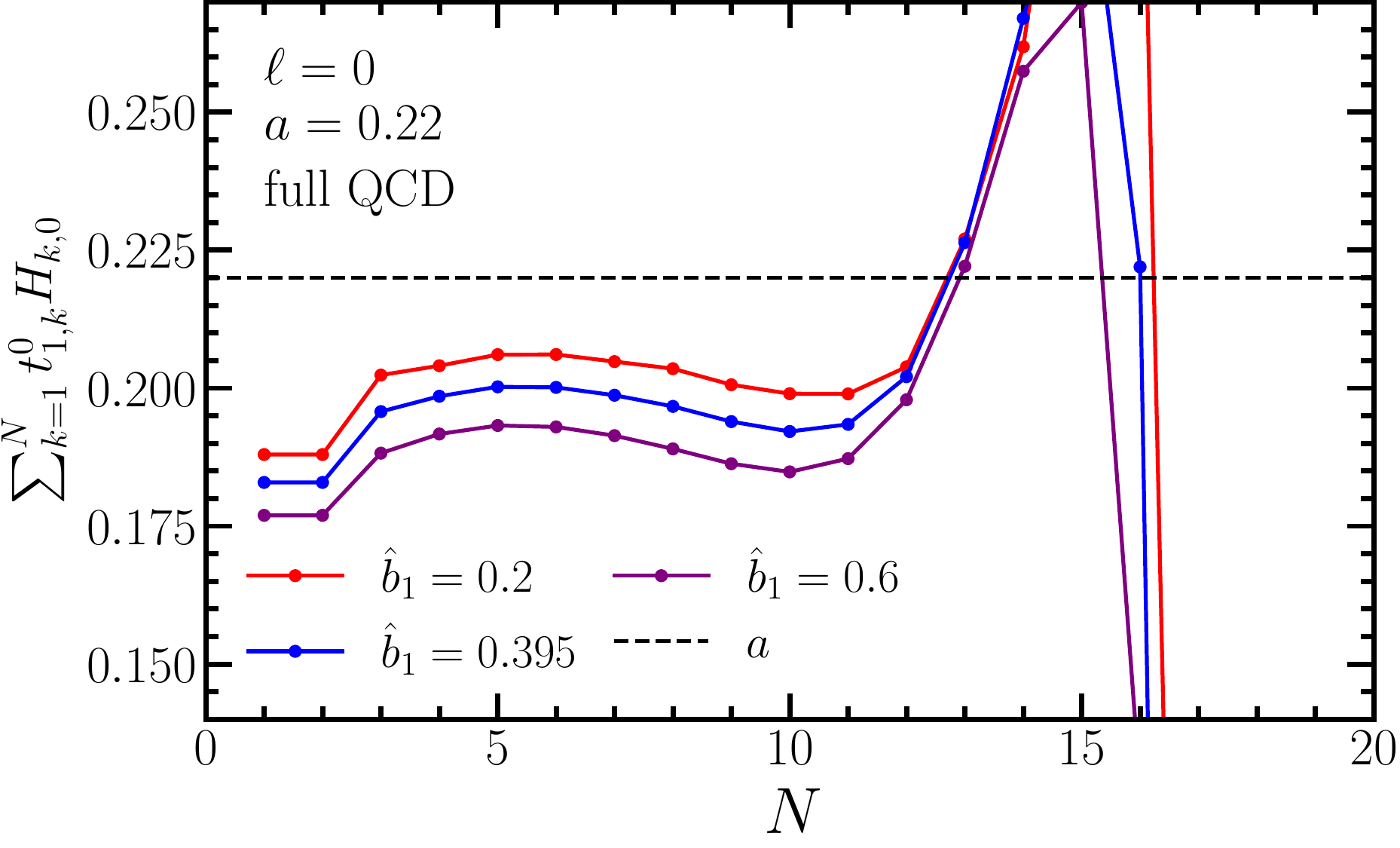}
\label{fig:anFullQCD1}}
\subfigure[]{\includegraphics[width=0.476\linewidth]{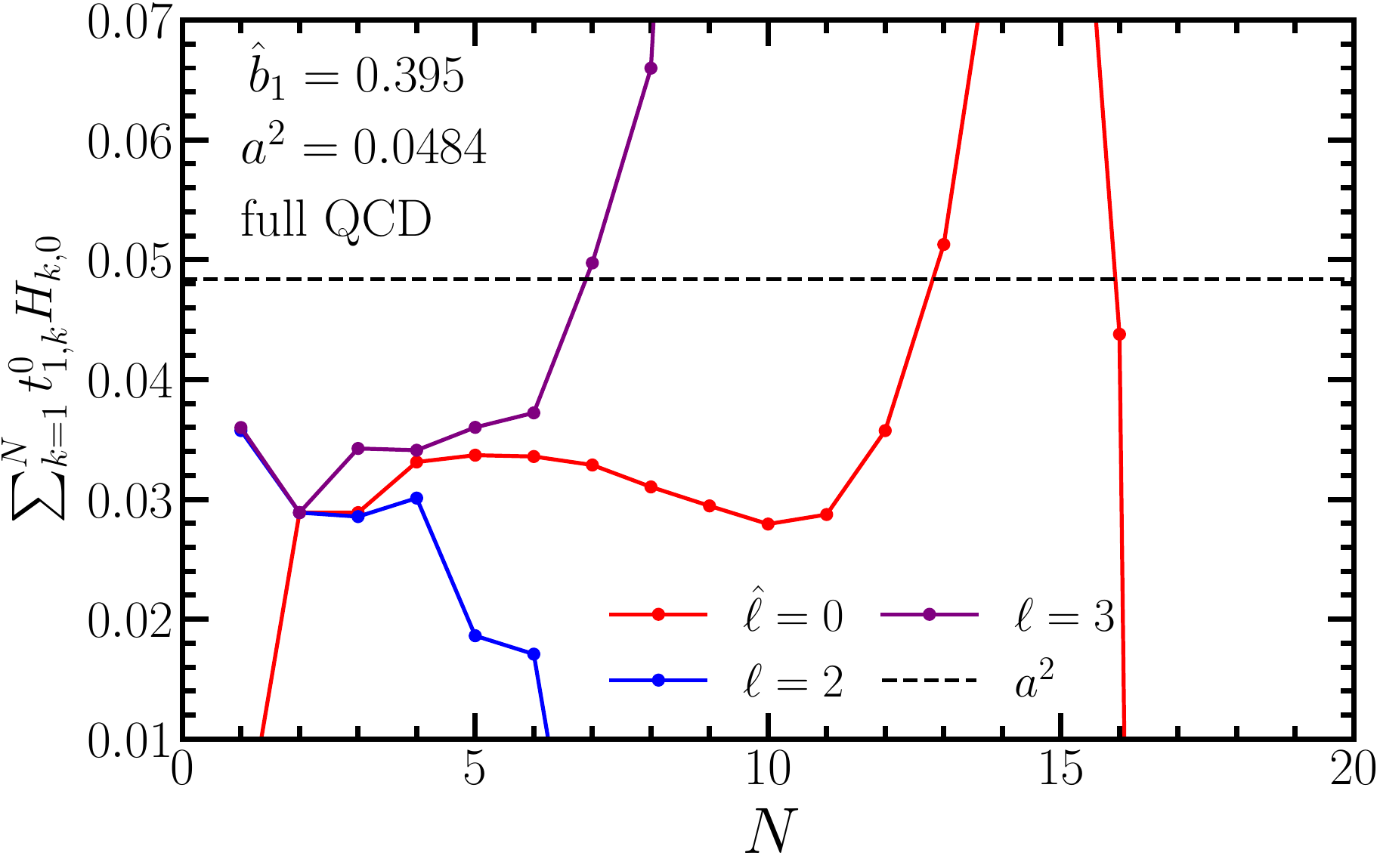}
\label{fig:anFullQCD2}}
\caption{Expansion \eqref{eq:aninHk} for full QCD using the C-scheme. The left panel shows the expansion of $a$ in terms of $H_{n,\ell}(a)$ for three different values of
$\hat b_1$: $0.2$ (red), $0.395$ (blue), and $0.6$ (purple). The right panel shows the expansion of $a^2$ in terms of $H_{n,\ell}(a)$ with $\hat b_1=0.395$ fixed, for
three different values of $\ell$: $0,1,$ and $3$ for red, blue, and purple, respectively.
Both panels use $a=0.22$.}
\label{fig:RunningSTFullQCD}
\end{figure*}

Finally, we also studied the large-order behavior of the $s^\ell_{n,k}$ coefficient for the series of the $H_{n,\ell}$ functions in powers of $a$ and the
$t^\ell_{n,k}$ coefficients for series of $a^n$ in terms of the $H_{n,\ell}$ functions, see Eqs.~\eqref{eq:Hninak} and \eqref{eq:aninHk}, respectively.
Our findings are briefly summarized as follows:
As for the large-$\beta_0$ approximation, the former series has a finite radius of convergence, while the latter does not converge for any value of $a$. Just like in the large-$\beta_0$ approximation, we find that the convergence radius arising from the $s^\ell_{n,k}$ coefficients for the expansion of the $H_{n,\ell}$ functions in powers $a^n$ agrees with the convergence radius obtained from the FOPT series coefficients and the expansion of the $a_\pm$ in $a^n$ powers, see Eq.~\eqref{eq:fitforconvradius}. For the $H_{n,0}$ this can also be seen directly from their analytic expressions given in Eqs.~\eqref{eq:H10fullQCDanalytic} and \eqref{eq:Hn0fullQCDanalytic} which are simple elementary functions of $a_\pm$. Their Taylor expansion in powers of $a$
yields a limit superior determined by the Taylor expansion of $a_\pm$.
For the $H_{n,\ell>0}$ the same property follows from the fact that the combination of the asymptotic expansions of the two functions $h_{n,\ell}$ in Eq.~\eqref{eq:HnellfullQCDanalytic} yields a convergent series, see the comment in Footnote~\ref{foot:gammaasy}.
Exemplarily, we have displayed the results for $|s^\ell_{n,k}|^{1/k}$ and $|t^\ell_{n,k}|^{1/k}$ as a function of $k$ for different combinations of $(n,\ell,\hat b_1)$ in Figs.~\ref{fig:SFullQCD} and ~\ref{fig:TFullQCD}, respectively.
For the $t^\ell_{n,k}$ we again find that their sequence in $k$ diverges factorially just as in the large-$\beta_0$ approximation for any $\ell$. The series of $a$ and $a^2$ in terms of the functions $H_{n,\ell}(a)$ is shown for different values of $(\ell,\hat{b}_1)$ in Figs.~\ref{fig:anFullQCD1} and \ref{fig:anFullQCD2} exhibiting again the apparent convergence at intermediate orders, the discrepancy to the actual value and the eventual divergence. The results show a similar behavior for any other value of $\ell$ and the divergence also arises for physical weight functions.

Overall, we find beyond all reasonable doubt that all features of the CIPT series expansion using the $H_{n,\ell}$ functions we found in the large-$\beta_0$ approximation are also present in full QCD. In particular, any finite FOPT series is in general divergent when transformed into a CIPT expansion.

\section{Summary and Conclusions}
\label{sec:conclusions}
In this article we have provided a detailed discussion on the mathematical aspects of CIPT that has been used as a main method to make perturbative predictions for $\tau$ hadronic spectral function moments. In contrast to the FOPT expansion in powers of the strong coupling at a particular renormalization scale, $[\alpha_s(s_0)]^n$, CIPT yields expansions in functions where an integration over the renormalization scale is carried out. So CIPT provides series in nontrivial functions of the strong coupling that differ fundamentally from the solutions of the RGE equation for the strong coupling which are used in FOPT. While previous discussions on the same subject were predominantly based on explicit calculations of the series to a particular order~\cite{Jamin:2005ip,Cvetic:2010ut,Caprini:2011ya} or the renormalon calculus, where different kinds of models for the Borel transform of the actual series were considered~\cite{Beneke:2008ad,Caprini:2009vf,Descotes-Genon:2010pyp,Beneke:2012vb,Abbas:2013usa,Hoang:2020mkw,Hoang:2021nlz}, we have carried out our analysis at a more basic level analysing directly the actual series using the elementary definitions and theorems on the (non)convergence of series and on asymptotic expansions.

For the most part of this article we have used the large-$\beta_0$ approximation, where all results can be obtained in an analytic way and all conclusions and considerations can be made in a mathematically rigorous way.
We have reconfirmed that CIPT provides a well-defined asymptotic expansion, like the common power series expansion. However, the CIPT expansion functions represent nonuniform asymptotic sequences due to zeros along the positive axis that approach $\alpha_s=0$ at large orders. We have proved that CIPT yields divergent series expansions without any region of convergence in cases where the OPE demands the existence of a region of convergence. This is in contrast to FOPT which indeed leads to series with a convergence region in these cases. We have found that the reason why the CIPT expansions frequently exhibits better series behavior than FOPT at low and intermediate orders is related to the zeros and because the CIPT expansion functions are bounded.

The most important finding of our analysis is that, while each CIPT expansion function has a finite region of convergence when written as a FOPT power series and represents an absolutely convergent series in this region, the inverse is not true. In other words, any FOPT power term $[\alpha_s(s_0)]^n$ yields a divergent series when expressed as a sum of CIPT expansion functions regardless of the value of $\alpha_s(s_0)$. There is a range of orders where the CIPT partial sum stabilizes before it diverges, but the truncated value of the CIPT series at these orders differs systematically from the value $[\alpha_s(s_0)]^n$.
In other words, a convergent FOPT series will in general diverge in CIPT, and using CIPT will in general yield a degradation of the convergence properties of a series and an apparent convergence may yield an unphysical value inconsistent with the OPE.
It is the latter property that makes phenomenological applications of CIPT dangerous.
This does not exclude the use of CIPT as a phenomenological expansion method, but it must be used with great care. A phenomenologically consistent way to apply CIPT was suggested in Refs.~\cite{Benitez-Rathgeb:2022yqb,Benitez-Rathgeb:2022hfj}.

Using models for expansion functions forming asymptotic sequences, we have shown that this property appears to be related to the zeros of the CIPT expansion functions which approach $\alpha_s=0$ at high orders. It is kind of ironic that these zeros appear to be harmless at first sight and even beneficial as they suppress the numerical size of the CIPT expansion functions. We have provided numerical evidence that all our findings obtained in the \mbox{large-$\beta_0$} approximation are also valid in full QCD. The primary use of our findings is that they can be used as a tool to also test the consistency of other expansion methods, where an integration over the renormalization scale is carried out.

\vspace*{0.8cm} {\bf Note Added:} During submission of this work we received Ref.~\cite{Golterman:2023oml} where the concept of the ``asymptotic separation'' put forward in Refs.~\cite{Hoang:2020mkw,Hoang:2021nlz} was confirmed based on an alternative approach.

\subsection*{Acknowledgements}
This work has been supported by the MECD Grant No.\ PID2019-105439GB-C22, the EU STRONG-2020 project under Program No.\
H2020-INFRAIA-2018-1, Grant Agreement No.\ 824093 and the COST Action No.\ CA16201 PARTICLEFACE.
N.G.G. has been supported by a JCyL scholarship funded by the regional government of Castilla y Le\'on and European Social Fund,
2017 call, and thanks the Particle Physics Group at the University of Vienna for hospitality while parts of this work were completed. We thank M.~Jamin and D.~Boito for carefully reading the manuscript and for their useful comments.

\appendix

\section{Review of Definitions and Theorems for Mathematical Series}
\label{sec:theorem}
In this appendix we collect a number definitions, theorems and corollaries on series and series of functions, as well as the concept of general asymptotic expansions, which we refer to in the main body of the article.
The statements can be found scattered throughout the mathematical literature in various formulations, but we quote them based on the presentations given in the classic text books by K.~Knopp~\cite{Knopp1990} (for Sec.~\ref{sec:seriestheorems}) and A.~Erd\'elyi~\cite{Erdelyi1955} (for Sec.~\ref{sec:asymptoticheorems}), where also explicit proofs can be found.

\subsection{Series and Series of Functions}
\label{sec:seriestheorems}

We start from the common definition concerning the convergence of an infinite series.

\begin{definition}
\label{th:defconvergence}
An infinite series $\sum_{i=1}^\infty c_i$ is called convergent if there exists a number $S$ such that for any $\epsilon>0$ there exists an integer $N(\epsilon)$ with the property that the sequence of partial sums $s_n=\sum_{i=1}^n c_i$ satisfies
$|s_n-S|<\epsilon$ for all $n\ge N(\epsilon)$. The number $S$ is then called the sum of the series, we say that the sum converges to $S$, and we write $S=\sum_{i=1}^\infty c_i$. If no such number $S$ exists the series is called divergent.
An infinite series $\sum_{i=1}^\infty c_i$ is called absolutely convergent if the series of absolute values $\sum_{i=1}^\infty |c_i|$ is convergent.
If $\sum_{i=1}^\infty c_i$ converges, but $\sum_{i=1}^\infty |c_i|$ does not, the series is called conditionally convergent.
\end{definition}

For considerations concerning the convergence of series we use the well-known and powerful root comparison test, which is formulated in the following root criterion:

\begin{theorem}[Cauchy's Root Criterion]
\label{th:root}
Let $\sum_{i=1}^\infty c_i$ be an infinite series. Let us denote $L\equiv\limsup_{n\to\infty}|c_n|^{1/n}$. Then the series is absolutely convergent if $L<1$, is divergent if $L>1$ or $L=\infty$. If $L=1$ the series may be either convergent or divergent.
\end{theorem}

The root test specifies the property of absolute convergence which is sufficient for our discussions. Since we apply our considerations for series in the strong coupling which has uncertainties, the particular case $L=1$ has no particular meaning for us, and will therefore not be discussed in any way. Since we do not only consider power series, but also series of functions we also introduce a generalization of the circle of convergence known for power series.

\begin{definition}[Interval of Convergence]
\label{th:intervalconvergence}
An interval $I$ is called interval of convergence of a series of functions $\sum_i^\infty f_i(x)$, if for any $x\in I$ all functions $f_i(x)$ are defined and the corresponding series converges.
\end{definition}

For infinite series of functions $\sum_{i=1}^\infty f_i(x)$ that converge to a function $F(x)$ the property of uniform convergence is important as linear operations on $F(x)$ can then typically be carried out by doing the operation on the individual functions $f_i(x)$ and summing afterwards.

\begin{definition}[Uniform Convergence]
\label{th:uniformconvergence}
A series of functions $\sum_{i=1}^\infty f_i(x)$ that converges to the function $F(x)$ in the interval $I$ is said to be uniformly convergent in an interval $I'\subseteq I$, if for any $\epsilon>0$, a number $N(\epsilon)$ independent of $x$ exists such that $|\sum_{i=1}^n f_n(x)-F(x)|<\epsilon$ for all $n\ge N(\epsilon)$ and all $x\in I'$.
\end{definition}

Weierstrass has formulated a simple bound criterion for uniform convergence. This theorem and the following corollary are important in our discussions as well.

\begin{theorem}[Weierstrass' Uniform Convergence Theorem]
\label{th:Weierstrassuniform}
If each of the functions $f_i(x)$ is defined and bounded in the interval $I$, i.e., $|f_i(x)|\leq \gamma_i$ for all $x\in I$, and if the series of positive terms $\sum_{i=1}^\infty \gamma_i$ converges, the series $\sum_{i=1}^\infty f_i(x)$ converges uniformly in $I$.
\end{theorem}

\begin{corollary}
\label{th:corollarypower}
A power series $\sum_{i=1}^\infty c_i\, x^i$ with interval of convergence $I$ converges uniformly in every subinterval $I^\prime \subset I$.
\end{corollary}

We finally quote the powerful and important Weierstrass theorem on the reordering of double series without referring to the absolute convergence of the double series.

\begin{theorem} [Weierstrass' Double Series Theorem]
\label{th:Weierstrassdouble}
Consider an infinite set of functions $f_i(x)$ that are analytic for $|x|<r$, so that the power expansions $f_i(x)=\sum_{k=0}^\infty a^{(i)}_k x^{k}$ exist and converge at least for $|x|<r$ for all $i$. Furthermore, consider a convergent series of these functions $F(x)=\sum_{i=0}^\infty f_i(x)$ that is uniformly convergent for $|x|\le \rho$ for every $\rho<r$, so that the series converges in particular everywhere within the interval
$|x|<r$ and defines the function $F(x)$ there.
Then, the infinite sums $A_k=\sum_{i=0}^\infty a^{(i)}_k$ are convergent and the infinite sum $\sum_{k=0}^\infty A_k x^{k}$ converges to $F(x)$ for $|x|<r$, so that
$F(x)=\sum_{i=0}^\infty (\sum_{k=0}^\infty a^{(i)}_k x^{k})= \sum_{k=0}^\infty (\sum_{i=0}^\infty a^{(i)}_k) x^{k}$ and is analytic for $|x|<r$.
\end{theorem}

\subsection{Order Symbols, Asymptotic Sequences and Asymptotic Expansions}
\label{sec:asymptoticheorems}

In gauge quantum field theories perturbative series are typically not convergent, but only asymptotic. We therefore also specify the basis of asymptotic expansions
collecting the relevant mathematical definitions.
In the following, $R$ is a domain in the complex plane and $x_0$ a point in $\bar{R}$, the closure of $R$. The functions $f(x)$, $g(x)$, and the
sequence of functions $\phi_1(x), \phi_2(x), \phi_3(x),\ldots$, are analytic in $R$. We furthermore abbreviate the latter sequence of functions as $\{\phi_n(x)\}$.
For the purpose of our considerations we always assume that $\{\phi_n(x)\}$ is an infinite set so that $n$ runs over all natural numbers.

We start by quoting the definition of the well known ${\cal O}$-relation and the lesser known ``little'' $o$-relation of two functions. While the ${\cal O}$-relation is a statement on two functions approaching a certain point at the same type of ``speed'' (e.g.\ linear or quadratic), the ``little'' $o$-relation states that one of the two function approaches that point much faster. It is important for the definition of asymptotic expansions.

\begin{definition}[${\cal O}$-relation]
\label{th:bigO}
It is said that $f={\cal O}(g)$ as $x\to x_0$ if there exists a constant $A>0$ and a neighborhood $U$ of $x_0$ so that $|f(x)|\le A|g(x)|$ for all $x\in R\cap U$.
\end{definition}

\begin{definition}[$o$-relation]
\label{th:littleo}
It is said that $f=o(g)$ as $x\to x_0$, if for any $\epsilon>0$ there exists a neighborhood $U_\epsilon$ of $x_0$ such that $|f(x)|<\epsilon |g(x)|$ for all $x\in R\cap U_\epsilon$. When $g(x\neq x_0)\neq 0$ in some neighborhood of $x_0$, the condition is equivalent to $\lim_{x\to x_0}f(x)/g(x)=0$.
\end{definition}

So e.g.\ $\sin(x)={\cal O}(x)$ as $x\to 0$, while $x^2=o(\sin(x))$ as well as $x^2={\cal O}(\sin(x))$ as $x\to 0$. Note that $f=o(g)$ always implies $f={\cal O}(g)$.

Asymptotic expansions approximate a function as a sum of other functions, which in order to be well-defined must form asymptotic sequences~\cite{Erdelyi1955}. The latter are specified by the following definitions,
where the special property of uniformity plays an important role in this article.

\begin{definition}[Asymptotic Sequence]
\label{th:asymptsequence}
An infinite sequence of functions $\{\phi_n(x)\}=\{\phi_1(x),\phi_2(x),\ldots\}$ is an asymptotic sequence as $x\to x_0$ in $R$ if $\phi_{n+1}=o(\phi_{n})$ for all $n$.
\end{definition}

\begin{definition}[Uniform Asymptotic Sequence]
\label{th:uniasymptsequence}
A sequence of functions $\{\phi_n(x)\}$ is an uniform asymptotic sequence as $x\to x_0$ in $R$ if $\phi_{n+1}=o(\phi_{n})$ uniformly in $n$.
This means that for any $\epsilon>0$ there exists a neighborhood $U_\epsilon$ of $x_0$ such that $|\phi_{n+1}(x)|<\epsilon |\phi_{n}(x)|$ for all $x\in R\cap U_\epsilon$ and all $n$.
\end{definition}

The requirement of an asymptotic sequence to be uniform excludes e.g.\ that the functions $\phi_n$ have singularities, but also zeros which approach $x_0$ for increasing $n$.
The sequence of power terms $\{x^n\}$ is a uniform asymptotic sequence as $x\to 0$ in $\mathbb{R}$ as one can easily check from the property that $x^{n+1}/x^n=x$ independently of $n$.

We are now ready to state the definition of an asymptotic expansion, where it is at this point not relevant whether the asymptotic sequence involved is uniform in $n$ or not.

\begin{definition}[Asymptotic Expansion]
\label{th:asymptexpansion}
The series $\sum_{n=1}^\infty c_n \phi_n(x)$, with $\{\phi_n(x)\}$ an asymptotic sequence as $x\to x_0$, is called an asymptotic expansion of $f(x)$ as $x\to x_0$ if
$f(x)-\sum_{n=1}^N c_n\phi_n(x)=o(\phi_N)$ as $x\to x_0$ for all $N\in\mathbb{N}$. At any finite order $N$ we write the asymptotic expansion in the form
$f(x) \stackrel{x\to x_0}{\asymp}\sum_{n=1}^N c_n\phi_n(x)$.
\end{definition}

From Definitions \ref{th:littleo} and \ref{th:asymptexpansion} one can derive the following corollary which states how to determine the coefficients $c_n$:
\begin{corollary}
\label{th:determinean}
The coefficients of the previously defined asymptotic expansion of $f(x)$ as $x\to x_0$ with respect to the asymptotic sequence $\{\phi_n(x)\}$ can be determined through the following recurrence formula: $c_n=\lim_{x\to x_0}[ f(x)-\sum_{i=1}^{n-1}c_i \phi_i(x)]/\phi_n(x)$. Furthermore, this implies that the coefficients are unique.
\end{corollary}

\subsection{Useful Asymptotic Expansions of Special Functions}
\label{sec:asymptoticformulae}

For the convenience of the reader we also quote some relations and (asymptotic) expansions concerning the incomplete
gamma function since they are used frequently in this article.
The incomplete gamma function is defined by the integral
\begin{equation}\label{eq:incGammaDef}
\Gamma(n,x)\equiv\int_x^\infty t^{n-1}e^{-t}\dd t \,,
\end{equation}
for any $n$ and $x$ where for complex $x$ the path is going horizontally to real $+\infty$, with a branch cut along the negative real $x$-axis if $n\notin\mathbb{N}$.
For $n\in\mathbb{N}$ it can also be written as a finite sum:
\begin{equation}\label{eq:incGammarel}
\Gamma(n,x) \stackrel{n\in\mathbb{N}}{=} e^{-x}\Gamma (n) \! \sum_{j = 0}^{n - 1} \frac{x^j}{\Gamma(j-1)}\,.
\end{equation}
It satisfies the recursive relation $\Gamma(n+1,x) = e^{-x} x^n + n \Gamma (n, x)$, such that for an
integer negative first argument it can be expressed in terms of $\Gamma(0, x)$ and a finite sum:
\begin{equation}\label{eq:recursive}
\Gamma(-n, x) \stackrel{n\in\mathbb{N}_0}{=} (- 1)^n
\biggl[ \frac{\Gamma (0, x)}{\Gamma(n+1)}
+ e^{- x} \sum_{j = 1}^n \frac{(-x)^{j - n - 1}}{(n+1-j)_j} \biggr].
\end{equation}
Note that this expression also provides an asymptotic expansion for large $n$ being $j=1$ the leading term.
The expansion of $\Gamma(n,z)$ as $|n|\to\infty$ for any finite $z$ and $n\neq 0,-1,-2\ldots$ reads
\begin{equation}
\label{eq:GammaAsy1}
\Gamma(n,z)- \Gamma(n) \stackrel{-n\notin\mathbb{N}_0}{=} -z^n e^{-z} \sum_{k=0}^\infty \frac{z^k}{(n)_{k+1}}\,.
\end{equation}
This expansion is even absolutely convergent for any finite $z$.
A particular case frequent in this work occurs when $z=\pm i\ell\pi$, with $\ell$ a positive integer. The following identify holds:
\begin{equation}
\label{eq:GammaAsy2}
\Gamma(n,-i\ell\pi,i\ell\pi) = \Gamma(n,-i\ell\pi)-\Gamma(n,i\ell\pi) \stackrel{-n\notin\mathbb{N}_0}{=} 2i(-1)^\ell\sum_{k=0}^\infty\frac{(\pi \ell)^{n+k}}{(n)_{k+1}}\sin\Bigl[(n+k)\frac{\pi}{2}\Bigr].
\end{equation}
The asymptotic expansion of $\Gamma(n,z)$ for large $|z|$ and ${\rm arg}(z) < 3\pi/2$ reads:
\begin{equation}
\label{eq:gammalargez}
\Gamma(n,z) \stackrel{|z|\to\infty}{\asymp} z^{n-1} e^{-z} \sum_{k=0}^\infty \frac{(n-k)_k}{z^k}\,.
\end{equation}
For $n\in\mathbb{N}$ the sum terminates at $k=n-1$ and Eq.~\eqref{eq:gammalargez} represents an exact identity.

The Bernoulli numbers $B_n$ arise in the summation of powers of integers and can be obtained from the Taylor coefficients of the generating function $x/(e^x-1)$
\begin{equation}
\frac{x}{e^x-1} = \sum_{n=0}^\infty\,\frac{B_n}{n!}x^n\,.
\end{equation}
For odd $n\ge 3$ the Bernoulli numbers are zero.
The leading term in the asymptotic expansion of the Bernoulli numbers as $n\to\infty$ reads
\begin{equation}
\label{eq:Bernoulliasy}
B_{n}\stackrel{n\to\infty}{\asymp}
=-\frac{2n!}{(2\pi)^n}\cos\Bigl(\frac{n\pi}{2}\Bigr),
\end{equation}
where the term $\cos(n\pi/2)$ accounts for $B_n$ being zero for large odd $n$ and the correct overall sign.
We also quote Stirling's formula $\Gamma (n + 1)=n! \stackrel{n\to\infty}{\asymp} \sqrt{2\pi n}(n/e)^n$, which is used several times.

\bibliography{sources}
\bibliographystyle{JHEP}

\end{document}